%% file: arxiv.tex
\documentclass{article}

    \PassOptionsToPackage{numbers, compress}{natbib}

\usepackage[preprint]{neurips_2024}

\usepackage[utf8]{inputenc} 
\usepackage[T1]{fontenc}    
\usepackage{hyperref}       
\usepackage{url}            
\usepackage{booktabs}       
\usepackage{amsfonts}       
\usepackage{nicefrac}       
\usepackage{microtype}      
\usepackage{xcolor}         
\usepackage{bm}
\usepackage{amsthm}
\usepackage{algorithm}
\usepackage{algorithmic}
\usepackage{amsmath}
\usepackage{amssymb}
\usepackage{mathtools}
\theoremstyle{plain}
\usepackage{multirow}
\usepackage{tabularx}
\usepackage{graphicx}
\usepackage{makecell}
\usepackage{color}
\usepackage{colortbl}
\usepackage{mathtools}
\usepackage[export]{adjustbox}
\usepackage{wrapfig}
\usepackage{enumitem}
\DeclareMathOperator*{\argmax}{arg\,max}

\newtheorem{theorem}{Theorem}

\title{Under-confidence Backdoors Are Resilient and Stealthy Backdoors}

\author{
Minlong Peng$^1$, Zidi Xiong$^1$, Quang H. Nguyen$^3$, Mingming Sun$^1$,  Khoa D. Doan$^3$, Ping Li$^2$
\\
Cognitive Computing Lab \\
Baidu Research \\
No.10 Xibeiwang East Road, Beijing 100193, China$^1$ \\
10900 NE 8th St. Bellevue, Washington 98004, USA$^2$
\\
College of Engineering and Computer Science, VinUniversity, Vietnam$^3$ \\
\{pengminlong,xiongzidi,sunmingming01,liping11\}@baidu.com\\
\{quang.nh, khoa.dd\}@vinuni.edu.vn
}

\begin{document}

\maketitle
\begin{abstract}
  Backdoor attacks aim to manipulate the victim model into producing specific outputs on any input injected with pre-designed triggers. Existing dirty-label backdoor attacks, despite showing high attack efficiency, suffer from the over-confidence problem: the victim models will behave peculiarly when the backdoor is implanted. Built upon this characteristic, existing defense algorithms are highly effective in mitigating the risks of these attacks. This work proposes a novel method to overcome this over-confidence problem in the existing attacks and accordingly, increase their stealthiness against the existing defense algorithms relying on this over-confidence characteristic. The rationale of our method is to reduce the backdoor effect in such a way that the victim model will only predict the poisoned input as the target class with a probability slightly greater than those of the other classes. This is achieved by re-designing the label-changing strategy of dirty-label backdoor attacks: the label of a poisoned input $\bm{x}$ will be changed to the target class with an instance-specific probability of $p_t(\bm{x})$, instead of 100\% as in previous methods. We empirically show that our method considerably improves the stealthiness of several representative backdoor attacks against the defense algorithms on four benchmark datasets in the backdoor domain. The results encourage backdoor researchers to develop defensive countermeasures to mitigate this type of attack.
\end{abstract}

\section{Introduction}
\vspace{-5pt}

Backdoor or trojan attack is a form of adversarial attack applied to the training process of models ~\cite{gu2017badnets}. It has become an increasing security threat, drawing many research interests in recent years \cite{chen2017targeted,nguyen2020input,doan2021lira,qi2021hidden,turner2019label,saha2020hidden,zhao2020bridging,souri2022sleeper}. 
To perform a backdoor attack, it usually first selects a few benign samples from the training set, inserts a pre-defined backdoor trigger to their inputs, and accordingly changes their labels to a designed target class. Then, it injects the poisoned samples back into the training set and provides the resulting data set to a victim for training the victim model. At inference time, given an input, the attack manipulates the victim model to predict the designed target class by injecting the corresponding pre-defined trigger to this input. 

In this field, dirty-label backdoor attacks (also called poison-label attacks in some related works), which consistently change the labels of poisoned samples to the target class, show great efficiency in attacking. They usually only need to poison 1\% or so of training data to achieve an almost 100\% attack success rate. However, at the same time, their corresponding victim models often show \textbf{severe over-confidence problems} on the backdoor samples: \textit{the victim model predicts the target class of a backdoor sample with a significantly higher posterior probability ($\gg 0.5$) than that of the second-best prediction}. Many defense algorithms are built on the resulting phenomenons of this characteristic. For example, STRIP~\cite{gao2019strip} observed that the label prediction of poisoned inputs would barely change over perturbations and accordingly proposed a perturbation-based strategy to {defend} the attack; Fine-Pruning~\cite{liu2018fine} observed that some neurons of the victim model would be activated only when the backdoor trigger appeared and accordingly proposed a {Fine-Pruning} method to disable the backdoor behavior. 

In this work, we propose a novel strategy to alleviate the over-confidence problem of existing dirty-label backdoor attacks and accordingly, improve their stealthiness against defensive algorithms that build on the over-confidence characteristic. The rationale of our strategy is to relax the correlation between the backdoor and the target label so that the poisoned input will be predicated as the target class as intended but the predicted probability of the target class is only slightly greater than those of the other classes. To achieve this goal and avoid the need 
for controlling the model training process, we change the label of a poisoned training sample to the target class with a specifically designed probability of $p_t(\bm{x})$, instead of $100\%$ as in the previous methods. It means the labels of some poisoned samples will not be changed to the target class. In addition, we give a theoretically-supported guidance for setting the value of $p_t(\bm{x})$. Besides, we show that it is flexible to control the predicting probabilities of poisoned inputs using our strategy. 

To evaluate the effectiveness of our method, we performed experiments on several existing, popular dirty-label backdoor attacks, including BadNet \cite{gu2017badnets}, SIG \cite{barni2019new}, and WaNet \cite{nguyen2021wanet}, and six prevalent defense algorithms, including Fine-Pruning, STRIP \cite{gao2019strip}, Neural-Cleanse \cite{wang2019neural}, NAD \cite{li2021neural}, Spectral Signature~\cite{tran2018spectral}, and SPECTRE~\cite{hayase2021spectre}. The empirical results show that our strategy can considerably improve the stealthiness of these existing attacks against these defense algorithms while keeping the merit of dirty-label backdoor attacks in attack efficiency.
In summary, our strategy brings the following advantages to existing dirty-label backdoor attacks:

\begin{itemize}[leftmargin=0.5cm]
    \item The connection between the backdoors and the target class is relaxed in our strategy. Consequently, our strategy can enhance the stealthiness of a dirty-label existing backdoor attack against defense algorithms built on the backdoor over-confidence characteristic of the victim model.

    \item The proposed method can manually control the degree of attack by controlling the number of applied attacks. Specifically, it can apply several different smoothed backdoors once at a time and manipulate the prediction probability of the target class by controlling the activated number of backdoor triggers at inference time. Usually, activating more backdoors leads to a higher prediction probability of the target class for a poisoned input. In contrast, in conventional dirty-label backdoor attacks, the predicting probabilities of the target class on poisoned inputs are always almost 100\%, making the attacks not flexible.
    
    \item The proposed method makes it harder for human inspectors to associate the backdoors with the target class since there are quite a few poisoned samples being injected with the triggers but not labeled as the target class.    
\end{itemize}

\vspace{-10pt}
\section{Background \& Related Work}
\vspace{-5pt}
\subsection{Backdoor Attack}
\vspace{-5pt}
Backdoor attacks aim at creating a trigger and associating it with a target class (or multiple triggers in the case of multiple target classes) so that the victim model will recognize any input containing the trigger as an instance of the target class, no matter what the form and true label of the input is. 

Formally, let $\bm{x}\in \mathcal{X}$ denote the input, $y \in \mathcal{Y}$ denote the output, $\mathcal{D}=\{(\bm{x}_1, y_1), \cdots, (\bm{x}_n, y_n)\}$ denote a clean training data set. To perform backdoor attacks, the attacker first {selects} a small number of benign samples from $\mathcal{D}$ and then, {injects} the pre-defined backdoor to each selected input and changes its label as follows:
\begin{equation*}
    \tilde{\bm{x}} = \mathcal{B}(\bm{x});
    \tilde{y} = T(y),
\end{equation*}
where $\mathcal{B}$ denotes the backdoor injection function, and $T$ denotes the relabel function. 
After that, the attacker merges the poisoned samples, ($\tilde{\bm{x}}$, $\tilde{y}$), with the rest of benign training data, and provides the 
resulting data set, $\Tilde{\mathcal{D}}$, to the victims. Without knowing the training data being poisoned, the victims train a model, $f_{vim}$, on $\tilde{\mathcal{D}}$, and accordingly provide a service based on $f_{vim}$. 
In the attack stage, the attacker can insert the trigger to any input $\bm{x}$ so that the victim model behaves as follows:
\begin{equation*}
    f_{vim}(\bm{x}) = y; f_{vim}(\mathcal{B}\left(\bm{x})\right) = T(y).
\end{equation*}


\vspace{-5pt}
\subsubsection{Dirty-label Backdoor Attack}
\vspace{-5pt}
This work studies the over-confidence problem of dirty-label backdoor attacks. Thus, we focus our discussion on dirty-label backdoor attacks in the following section.
dirty-label backdoor attacks change the label of the poisoned samples to the target class (in the all-to-one setting), i.e., $T(y) := c_t$ \cite{liu2017trojaning,liu2020reflection}. 
These attacks can be quite effective. They usually only need to poison 1\% or so of the training samples to achieve an almost 100\% attack success rate. 

As a pioneering work, {BadNet} \cite{gu2017badnets} used stickers and checkerboards as trigger patterns for image data. It replaces a fixed area of the input image with the pattern to poison the image, i.e., $\mathcal{B}(\bm{x}) = \bm{m}\odot\bm{x} + (1-\bm{m}) \odot \bm{\eta}$, where $\bm{m}$ is a binary mask matrix and $\bm{\eta}$ is an image with checkerboards in the corners. After that, many variants of backdoor attacks were introduced, most of which focused on decreasing the visibility of the backdoor trigger. For example, Blend \cite{chen2017targeted} used a normal image as the trigger pattern and mixed the trigger pattern with the benign input to inject the backdoor.
SIG \cite{barni2019new} proposed to add a horizontal sinusoidal signal to the benign image as the backdoor for data poisoning.
{WaNet} \cite{nguyen2021wanet} used a small and smooth warping field as the trigger pattern. Another work~\cite{sun2020natural} proposed to inject special characters, words, and phrases into text data as trigger patterns,
while Hidden Killer~\cite{qi2021hidden} proposed to poison the textual inputs by making them follow a specific syntactic rule, which rarely occurs in benign data. 
{ISSBA} \cite{li2021invisible} proposed an instance-specific attack strategy that generates a specific trigger pattern for each of the poisoned inputs. For this purpose, it trains an encoder-decoder model. The encoder takes the concatenated representation of an input image and an attacker-defined code vector as input and reconstructs the input image. The decoder takes the reconstructed image as input and reconstructs the code vector. The trained encoder is then used to poison the data. 

\vspace{-5pt}
\subsubsection{Over-confidence of Dirty-label Backdoor Attack}
\vspace{-5pt}
Dirty-label backdoor attacks show great attack efficiency. However, their victim models often show severe over-confidence on the backdoor samples, which makes them easily exposed to defense algorithms. For example, some neurons of the victim model may be abnormally large when the backdoor is injected \cite{xu2020defending}, and the variation of the label prediction may be extremely small to noise for the poisoned input \cite{gao2019strip}. Here, we give our analysis.

For analysis convenience, suppose that $\mathcal{B}(\bm{x}) =: (\bm{x}, \bm{z})$, where $\bm{z}$ denotes the additional feature introduced by the backdoor to the benign input $\bm{x}$. Then, we have:
\begin{align}
    p(y|\bm{x}, \bm{z}) = \frac{p(\bm{x})p(y|\bm{x})p(\bm{z}|\bm{x}, y)}{p(\bm{x}, \bm{z})}.
\end{align}
Because the examples are randomly sampled to poison, $\bm{x}$ is independent of $\bm{z}$. Thus, 
\begin{align}
    p(y|\bm{x}, \bm{z}) &= \frac{p(y|\bm{x})p(\bm{z}|y)}{p(\bm{z})} = \frac{p(y|\bm{x})p(\bm{z}, y)}{p(\bm{z})p(y)} 
    =p(y|\bm{z}) \cdot \frac{p(y|\bm{x})}{p({y})}.
\end{align}
When the input, $\bm{x}$, is sampled from non-target classes (this is often the case in dirty-label backdoor attacks), $p(y=c_t|\bm{x})$ will be small, and accordingly, $p(y=c_t|\bm{x})/p(y=c_t)$ will be small. Then, to achieve a small training loss on $(\mathcal{B}(\bm{x}), c_t)$, the victim model has to increase the value of $p(y|\bm{z})$ so that to increase the value of $p(y=c_t|\bm{x}, \bm{z}$). This will result in the victim model becoming over-confident over $\bm{z}$ to overcome the influence of $\bm{x}$ on the label prediction of $\mathcal{B}(\bm{x}) = (\bm{x}, \bm{z})$. Of course, in real cases, it may be not easy to separate $\bm{z}$ from $\bm{x}$ for poisoned inputs, and the representation of $\bm{x}$, and accordingly $p(y|\bm{x})$, will be affected by $\bm{z}$ in the victim model. However, the above analysis is still instructive. In addition, the analysis indicates another direction for alleviating the over-confidence problem of dirty-label backdoor attacks we do not study in this work, i.e., poisoning samples that are close to the target class boundary so that $p(y=c_t|\bm{x})$ will not be very small. We leave this study in future work.

\vspace{-5pt}
\subsection{Backdoor Attack Defense}
\vspace{-5pt}

Backdoor defenses aim to verify or mitigate the provided model before deployment. Several categories of defense strategies have been developed in recent years to counter backdoor attacks, including explicit backdoor mining \cite{chen2019deepinspect,wang2019neural,liu2019abs,zeng2021rethinking}, poisoned input detection \cite{gao2019strip,xu2020defending,yang2021rap,li2021backdoor,li2021anti}, and model mitigation \cite{liu2018fine,zhao2020bridging,huang2022backdoor} approaches. Many of these approaches build on the idea that the attacked model strongly associates the trigger to the target class and will, therefore, act over-confidently when the backdoor input is presented. 

For instance, previous works~\cite{liu2018fine, xu2020defending} observed that some neurons of the victim model would only be activated when the trigger pattern appeared. Therefore, they proposed disabling the backdoor behaviour by eliminating neurons that were dormant on benign inputs. With similar motivation, NAD~\cite{li2021neural} finetuned the victim model with a few benign samples, then aligned the neurons of the victim model with those of the finetuned model through a neural attention distillation process to perform the defense. 
Neural Cleanse \cite{wang2019neural} assumes that the backdoor trigger is patch-based and can easily change the prediction label of any input to the target class. For each class, it optimizes a patch pattern that makes the model classify inputs with the pattern as the class. If any class yields a significantly smaller pattern, Neural Cleanse considers it as a potential backdoor.
STRIP \cite{gao2019strip} perturbs the given input image and determines the presence of a backdoor according to the entropy of the predictions of the perturbed images. Due to the over-confidence of the victim model over the backdoor, the entropy of a poisoned input should be smaller than that of a benign input. On the other hand, outlier removal defenses~\cite{tran2018spectral, chen2018activationclustering, hayase2021spectre} are proposed to filter training samples that are likely to be poisoned and train the model on the new dataset.

In expectation, if we can alleviate the over-confidence of the victim model over the backdoor, the effectiveness of these defense algorithms will degrade. For instance, the activation of the backdoors will not be abnormally large so that the attack can escape the defense of neuron-pruning-based methods and NAD;
the backdoor may result in large optimization loss for the target class so that the attack can escape the delectation of Neural Cleanse; the victim model will not predict any class over-confidently so that the attack can escape the detection of STRIP. 

Of course, we should note that our method increases the stealthiness of the attacks by mitigating the backdoor over-confidence problem of the victim model. \textit{For the defense algorithms not built on the over-confidence characteristic of the victim model \cite{tran2018spectral,zeng2021rethinking,huang2022backdoor}, our method may not work.} However, in this case, we can combine our strategy with other techniques specifically designed for those defense algorithms. This is often necessary since the victims may run multiple algorithms for the defense. 
\vspace{-5pt}
\subsection{Label Smoothing}
\vspace{-5pt}
Label smoothing \cite{szegedy2016rethinking} is a widely used ``trick" to alleviate model over-confidence and improve model generalization. Specifically, label smoothing replaces one-hot encoded label vector $\bm{y}$ with a mixture of $\bm{y}$ and the uniform distribution:
\begin{equation*}
    \bm{\hat{y}}_{i} = (1-\alpha)\bm{y}_i + \frac{\alpha}{C},
\end{equation*}
where $C$ denotes the class number, and $\alpha$ is a hyper-parameter that determines the degree of smoothing. If $\alpha = 0$, we obtain the original one-hot encoded $\bm{y}$. If $\alpha = 1$, we get the uniform distribution. It has shown that label smoothing results in better model calibration and prevents over-confident predictions. 
\vspace{-10pt}
\section{Methodology}
\vspace{-5pt}
\noindent \textbf{Threat Model.} In this work, we follow the widely applied training data poisoning setting. The attacker can arbitrarily modify the training data but cannot change other training components (e.g., model structure and training loss). This is the scenario where the users adopt third-party collected data for model training.

The over-confidence problem of dirty-label backdoor attacks originates from the learning on the poisoned samples, which can be seen as noised samples in label noise learning \cite{frenay2013classification}. Thus, intrinsically, many existing techniques for addressing label noise can be applied to alleviate the over-confidence problem of backdoor attacks \textit{if we have access to the learning process of the model}. For example, we may change the cross-entropy loss to the more robust mean square error loss, or we may apply label-smoothing to the cross-entropy loss. \textit{The problem is that we have only access to the training data, thus most of these techniques are not applicable.} With this consideration and motivated by the success of the label-smoothing technique, we propose a novel data-level smoothing strategy to alleviate the over-confidence problem of backdoor attacks. \textbf{This strategy is not limited to any backdoor injection function, $\mathcal{B}(\cdot)$, and can be {generally} applied to all existing dirty-label attacks. }

\subsection{Overview}

In the proposed strategy, the sample is poisoned as follows:
\begin{equation} \label{eq:poison}
    (\bm{x}, y) \rightarrow \left\{\begin{matrix*}[l]
    (\tilde{\bm{x}}, c_t) & \text{with probability } p_t(\bm{x}); \\
    (\tilde{\bm{x}}, y) & \text{with probability } 1-p_t(\bm{x}),
    \end{matrix*}\right.
\end{equation}
where $p_t(\bm{x})$ denotes the probability to change the label of the benign sample to the \underline{t}arget class. Note that the conventional poisoned-label backdoor attack strategy and clean-label backdoor attack strategy are special cases of the above paradigm, with $p_t(\bm{x}) \equiv 1$ and $p_t(\bm{x}) \equiv 0$, respectively. To set the value of $p_t(\bm{x})$, we train a clean model, $f_{clean}$, on the benign training data set $\mathcal{D}$, then obtain the predicting probabilities, $\alpha(\bm{x})$ and $\lambda(\bm{x})$, on the target class and the predicted class of $\bm{x}$ given by $f_{clean}$, respectively, and finally obtain $p_t(\bm{x})$ by: $p_t(\bm{x}) = \frac{\beta(\bm{x}) - \alpha(\bm{x})}{1-\alpha(\bm{x})}$ with $\beta(\bm{x}) = \text{min}(\lambda(\bm{x})+0.1, 0.6)$.

\subsection{Address Backdoor Over-confidence with Data-Level Label Smoothing}
Recall that the attack is successful on a poisoned input, $\tilde{\bm{x}}$, if only:
\begin{equation}
    c_t = \argmax_{c} p \left(y=c|\tilde{\bm{x}}\right).
\end{equation}
Here, $p\left(y=c|\tilde{\bm{x}}\right)$ denotes the predicting probability of class $c$ given by the victim model $f_{vim}$ on $\tilde{\bm{x}}$. It does not require $p\left(y=c_t|\tilde{\bm{x}}\right) \rightarrow 1$. Our strategy builds on this principle and aims to set $p_t(\bm{x})$ in Eq. (\ref{eq:poison}) to the value that just make $c_t = \argmax_{c} p\left(y=c|\tilde{\bm{x}}\right)$ hold. A naive solution is to consistently set $p_t(\bm{x})$ to a value just greater than 0.5, e.g., 0.6. This solution achieves the smoothing purpose but is not very efficient. The corresponding values of $p_t(\bm{x})$ for inputs close to the classification boundary of the target class should be smaller than those of inputs far from the target class classification boundary. Actually, as revealed by the following theorem, for achieving the smoothing purpose, it only needs to set $p_t(\bm{x}) \approx 0.2$ and $p_t(\bm{x}) \approx 0.5$, for inputs close to and inputs far from the target class classification boundary, respectively.

\begin{theorem} \label{theorem1}
Let $\alpha(\bm{x}) :=p(y=c_t|\bm{x}; \mathcal{D})$ denotes the probability of $\bm{x}$ belonging to the target class, $\hat{\beta}({\bm{x}}) := p(y=c_t|\tilde{\bm{x}}; \tilde{\mathcal{D}})$ denotes the predicting probability of $\tilde{\bm{x}}$ belonging to the target class by the victim model, then we have:
\begin{equation} \label{eq:expect_beta}
    {\beta}({\bm{x}}) \equiv \mathbb{E}[\hat{\beta}({\bm{x}})] = \alpha(\bm{x}) + (1-\alpha(\bm{x}))p_t(\bm{x}).
\end{equation}
\end{theorem}

\begin{proof}
See the appendix.
\end{proof}

In order to set the value of $p_t(\bm{x}_i)$, we reformulate Eq. (\ref{eq:expect_beta}) as follows:
\begin{equation}
    p_t(\bm{x}) = \frac{{\beta}({\bm{x}})-\alpha(\bm{x})}{1-\alpha(\bm{x})}.
\end{equation}
For an intuitive understanding of the above formulation, consider the binary classification problem. To perform smoothing as much as possible on the premise of achieving a successful attack, we set ${\beta}({\bm{x}})$ just slightly greater than 0.5. In addition, note that $\alpha(\bm{x}) < 0.5$ for non-target class samples. Thus, $p_t(\bm{x})$ decreases by the value of $\alpha(\bm{x})$. That is, the closer the selected sample is to the target class, the lower the probability that we change its label to the target class is. 

In practice, we train a classifier, $f_{clean}$, using the clean training set to estimate the value of $\alpha(\bm{x})$ and set the value of $\beta(\tilde{x})$. Specifically, let $\hat{p}$ denote the modeled probability by $f_{clean}$. We estimate $\alpha(\bm{x})$ by:
\begin{equation}
    \alpha(\bm{x}) \leftarrow \hat{p}(y=c_t|\bm{x}),
\end{equation}
and set $\beta(\tilde{\bm{x}})$ by:
\begin{equation} \label{eq:beta}
    \beta({\bm{x}}) \leftarrow \min(\max_c \hat{p}(y=c|\bm{x}) + 0.1, 0.6).
\end{equation}
Once $\alpha(\bm{x})$ and $\beta(\tilde{\bm{x}})$ have been set, $p_t(\bm{x})$ is accordingly set. Notice that $\beta(\tilde{\bm{x}})$ is always greater than 0.5 in this practice for the binary classification task and may be smaller than $0.5$ in multi-class classification tasks. 

\subsection{Increase Predicting Confidence by Backdoor Stacking}

In conventional dirty-label backdoor attacks, the label prediction on a poisoned input is highly confident, and the predicting probability of the target class is almost 100\%. To alleviate the overconfidence of attacks, our strategy scarifies such confidence, encouraging the predicting probability of the target class to be $\beta({\tilde{\bm{x}}}) < 100\%$. Thus, a natural question arises: \textit{Can we increase the label prediction confidence on poisoned inputs while not aggravating backdoor over-confidence?}

We propose to address  this question by splitting the poisoned set into multiple subsets, each of which applies a unique backdoor:

\begin{theorem} \label{theorem2}
Suppose $k$ different backdoors are applied, which share the same $p_t(\cdot)$ function but have different backdoor injection functions $\mathcal{B}_1, \cdots, \mathcal{B}_k$, and select the poisoned samples independently. Let $\bm{\zeta}(\bm{x}) := \mathcal{B}_k(_{\cdots}\mathcal{B}_1(\bm{x})_{\cdots})$, $\hat{\beta}(\bm{x}) := p(y=c_t|\mathcal{B}_i(\bm{x}))$,
$\hat{\beta}^\prime(\bm{x}) := p(y=c_t|\bm{\zeta}(\bm{x}_i))$. Then, we have:
\begin{equation}
\begin{split}
    &{\beta}^\prime(\bm{x}) > \beta(\bm{x}), \\
    {\beta}(\bm{x}) \equiv \mathbb{E}[\hat{\beta}(\bm{x})] & = \alpha(\bm{x}) + (1-\alpha(\bm{x}))p_t(\bm{x}), \\
     {\beta}^\prime(\bm{x}) \equiv \mathbb{E}[\hat{\beta}^\prime(\bm{x})] &= \alpha(\bm{x}) + (1-\alpha(\bm{x}))\left[1-(1-p_t(\bm{x}))^k\right].
\end{split}
\end{equation}
\end{theorem}
\begin{proof}
See the appendix.
\end{proof}

Suppose that we design to poison $kn$ training samples to perform the attack. According to the above theorem, compared with the method that poisons all the $kn$ samples with a unique backdoor, it is more effective to apply $k$ different backdoors, each of which poisons $n$ training samples. Thus, in practice, we apply multiple different backdoors simultaneously on $\mathcal{D}$ (\textit{the total number of poisoned samples is controlled to be the same as that of the single-backdoor-based method for fair comparisons}). For each backdoor, we {independently} select the samples to poison at the data poisoning stage. In the attack stage, we activate one or multiple backdoors to perform the attack. 

\section{Experiments}

\begin{table*}[ht!]
    \centering
    \footnotesize
    \setlength{\tabcolsep}{4pt}
    \caption{Clean Accuracy (CleanAcc) $\times 100$ and ASR $\times 100$ without any defense. ``w/o" means not applying our strategy and ``w" means applying our strategy. 
    }
    \begin{tabular}{l|cccccccccccc@{}}
    \Xhline{2\arrayrulewidth}
    \multirow{3}{*}{Dataset} & \multicolumn{4}{c}{BadNet} & \multicolumn{4}{c}{SIG} & \multicolumn{4}{c}{WaNet} \\ \cline{2-5} \cline{6-9} \cline{10-13} 
    & \multicolumn{2}{c}{CleanAcc} & \multicolumn{2}{c}{ASR} & \multicolumn{2}{c}{CleanAcc} & \multicolumn{2}{c}{ASR} & \multicolumn{2}{c}{CleanAcc} & \multicolumn{2}{c}{ASR} \\ 
    & w/o & w & w/o & w & w/o & w & w/o & w & w/o & w & w/o & w \\ \cline{1-1} \cline{2-3} \cline{4-5} \cline{6-7} \cline{8-9} \cline{10-11} \cline{12-13}
    MNIST    & 99.26 & 99.21  & 100 & 100 & 99.22 & 99.20 & 100 & 100 & 99.25 &99.31 &98.95 &100  \\
    CIFAR10  & 93.57  & 93.60  & 99.92  &98.79  & 94.40 & 94.21 &99.96 & 94.32 & 94.05 &94.31 & 99.37 &100\\
    GTSRB & 98.12 & 97.93 & 100 & 100 &98.13 &97.55 &99.36 &98.56  & 98.91 &98.50 &99.57 &99.64 \\
    CelebA & 78.98 & 79.43 & 100  & 99.98 & 79.16 & 79.39 & 99.87 & 96.67 &79.53 &79.03 &99.87 &99.77\\
    \Xhline{2\arrayrulewidth}
    \end{tabular}
    
    \label{tab:static_attack_results}
\end{table*}
\subsection{Experiment Setup}

\noindent\textbf{Datasets.} Following the previous work, we performed experiments on four widely-used datasets: MNIST \cite{lecun1998gradient}, CIFAR-10 \cite{krizhevsky2009learning}, GTSRB \cite{stallkamp2012man}, and CelebA \cite{liu2015deep}. For the implementation of $f_{clean}$ and $f_{vim}$, we follow most of the setting of WaNet \cite{nguyen2021wanet} and considered a mixed of popular models: VGG11 and Pre-activation Resnet-18 \cite{he2016deep} for CIFAR-10 and GTSRB and Resnet-18 
for CelebA. For MNIST, we employed the same CNN model used by WaNet. 

Detailed information is provided in the appendix.

\noindent\textbf{Backdoor Injection Functions.}
The proposed strategy is not limited to any specific backdoor injection function, $\mathcal{B}(\cdot)$. Thus, we performed experiments with the following popular backdoor injection functions:
\begin{itemize}[leftmargin=0.5cm]
    \item \textbf{BadNet} \cite{gu2017badnets}: It injects a checkerboard on either top-left, top-right, bottom-left, or bottom-right corner of the input image. We implemented four different backdoor injection functions from this template with each function injecting the backdoor trigger on a unique corner. The backdoor of this kind of attack is small and relatively less stealthy. 
    
    \item \textbf{SIG} \cite{barni2019new}: It adds a horizontal sinusoidal signal to the benign image as the backdoor, which is defined by $v(i, j) = \Delta\sin(2 \pi j f/m), 1 \leq j \leq m, 1 \leq i \leq l$, where $m$ and $l$ denotes is the number of columns and rows of the image, and $f$ is a frequency term. We implemented four backdoor injection functions from this template by setting different values of $f=[2, 4, 8, 16]$ with $\Delta=0.05$. The backdoor of this kind of attack is relatively larger and less stealthy.
    
    \item \textbf{WaNet} \cite{nguyen2021wanet}: It uses a small and smooth warping field to generate the backdoor images. We implemented four backdoor injection functions from this template by using different warping fields. The backdoor of this kind of attack is significantly more stealthy.
\end{itemize}


\begin{wraptable}{r}{0.5 \textwidth}
    \vspace{-10pt}
    \setlength{\tabcolsep}{2pt}
    \footnotesize
    \centering
    \caption{ASR $\times$ 100. 4\%$\times$1: the attack applying one backdoor and poisoning 4\% of training samples. 1\%$\times$4: the attack applies 4 different backdoors with each poisoning 1\% of training samples.}
    \begin{tabular}
    {l|cccccc@{}}
    \Xhline{2\arrayrulewidth}
    \multirow{2}{*}{Dataset} & \multicolumn{2}{c}{BadNet} & \multicolumn{2}{c}{SIG} & \multicolumn{2}{c}{WaNet} \\ 
    & 4\%$\times$1 & 1\%$\times$4 & 4\%$\times$1 & 1\%$\times$4 & 4\%$\times$1 & 1\%$\times$4 \\ \cline{1-1} \cline{2-3} \cline{4-5} \cline{6-7} 
    MNIST   & 90.31 & 100 & 91.40 & 100 & 84.54 & 100  \\
    CIFAR10 & 81.21 & 98.79 & 84.65  & 94.32 & 31.43 & 100 \\
    GTSRB   & 81.05 & 100 & 87.04 & 98.56 & 34.90 & 99.64 \\
    CelebA  & 81.31 & 99.98 & 91.95  & 96.67 & 62.78 & 99.77\\
    \Xhline{2\arrayrulewidth}
    \end{tabular}

    \label{tab:attack_stacking}
\end{wraptable}

\noindent\textbf{Implementation.}
For each backdoor injection template, we implemented four backdoor injection functions. For our method, we injected the four backdoors one at a time to perform data poisoning, with each backdoor injected to \textbf{1\%} of training samples. In the attack stage, we manipulated the victim model by activating different numbers of backdoors and studied the attack performance. For the baseline (without our strategy) on each template, we selected the most effective backdoor injection function from the four candidates of the template and poisoned \textbf{4\%} of training samples so that the baseline attack and the attack with our strategy poison the same number of samples. For WaNet, we did not use the noise mode. Please refer to the source code in the appendix for additional implementation detail.

\subsection{Attack Experiments}

\subsubsection{Attack Effectiveness} 
Table \ref{tab:static_attack_results} shows the performance of the victim models on the clean and poisoned test sets. For attacks by our strategy, we activated all four backdoors in the attack stage to obtain the reported performance in this table. As can be seen from the table, on the clean data set, the victim models of the smooth attacks perform quite similar to those of the hard attacks, with accuracy near 100\% on MNIST/GTSRB, 94\% on CIFAR10, and 79\% on CelebA. On the poisoned data set, the smooth attacks also perform similarly to their hard counterparts, achieving above 90\% attack success rate on all datasets. These results show that our label-smoothing strategy would not degrade the attack effectiveness. 

\subsubsection{ASRs at Different Activated-Backdoor Numbers}
\begin{figure*}[t!]
    \centering
    \includegraphics[width=1.0\linewidth]{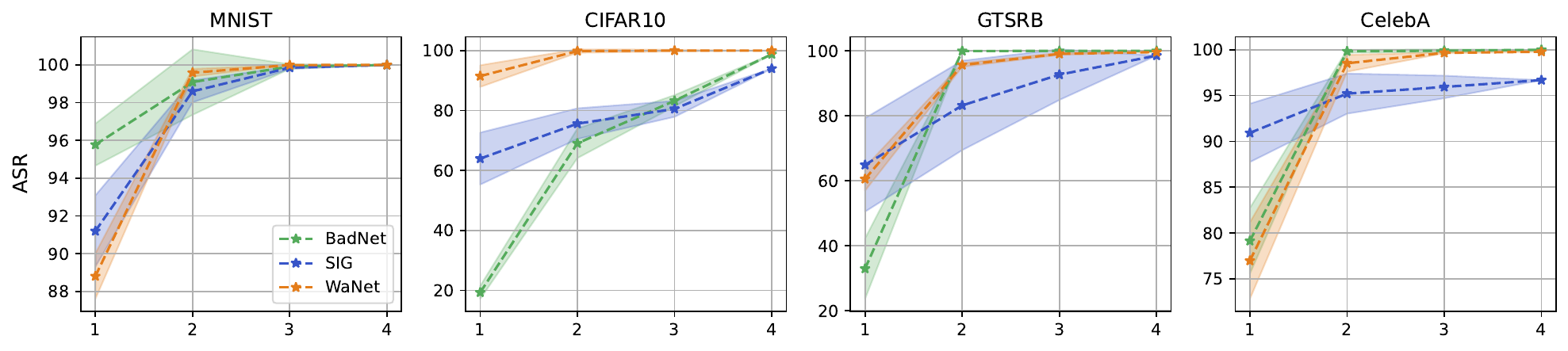}
    \vspace{-20pt}
    \caption{Average ASR ($\pm$ standard value) by the number of activated backdoors.}
    \label{fig:asr_by_attack_num}
    \vspace{-10pt}
\end{figure*}

Here, we first study the influence of the activated backdoor number on the attack performance of our method at inference time (at training time, 4 backdoors are applied). Figure \ref{fig:asr_by_attack_num} shows the results. As can be seen from the figure, the attack performance (ASR) of our method generally increases as the number of activated backdoors increases, and it usually achieves an acceptable attack performance when two backdoors are activated in the attack stage. 

We further study if it is more effective to use $k$ different backdoors to poison the data, where each backdoor poisons $n$ samples, than poisoning $kn$ samples using a single backdoor as revealed in theorem \ref{theorem2}. Specifically, for this study, we compare the performance of the attack that poisons 4\% of the training samples with the attack that apply four different backdoors, each of which poisons 1\% of the training samples. Both of these two attacks utilize our strategy. Table \ref{tab:attack_stacking} shows the results of this study. From the table, we can see that the attack performance of the single-backdoor attack is much worse than that of the multiple-backdoor attack although they poison the same number of training samples. This verifies our deduction in theorem \ref{theorem2}. 
\begin{figure*}[t!]
    \centering
    \includegraphics[width=1.0\linewidth]{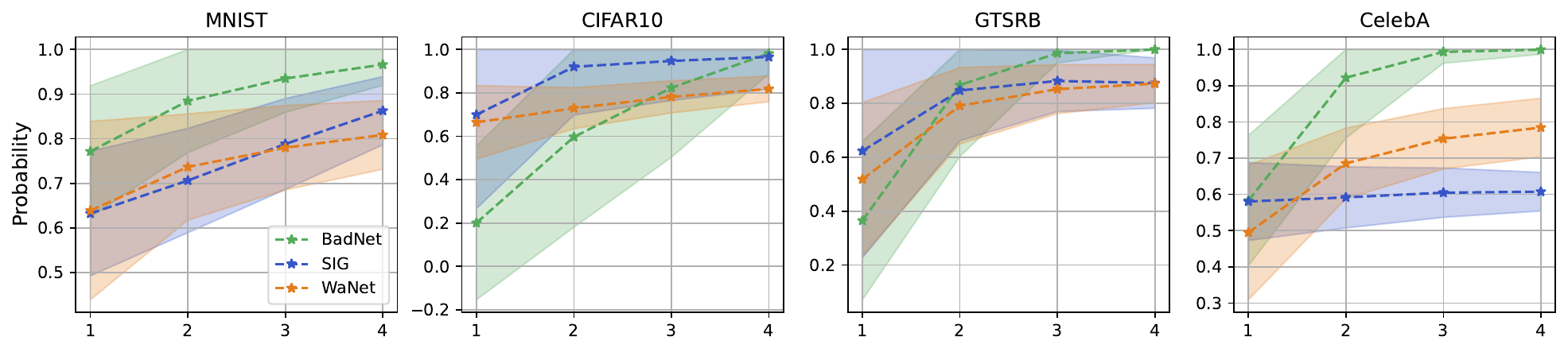}
    \vspace{-20pt}
    \caption{Average predicting probability ($\pm$ standard value) of the target class by the number of activated backdoors.}
    \label{fig:probs_by_attack_num}
    \vspace{-10pt}
\end{figure*}
\begin{table*}[t!]
    \centering
    \footnotesize
    \caption{ASR $\times$ 100. A larger value indicates better stealthiness against Fine-Pruning and NAD.}
    \resizebox{1.0\linewidth}{!}{
    \begin{tabular}{l|cccccccccccc@{}}
    \Xhline{2\arrayrulewidth}
    \multirow{3}{*}{Dataset} & \multicolumn{4}{c}{BadNet} & \multicolumn{4}{c}{SIG} & \multicolumn{4}{c}{WaNet} \\ \cline{2-5} \cline{6-9} \cline{10-13} 
    & \multicolumn{2}{c}{Fine-Pruning} & \multicolumn{2}{c}{NAD} & \multicolumn{2}{c}{Fine-Pruning} & \multicolumn{2}{c}{NAD} & \multicolumn{2}{c}{Fine-Pruning} & \multicolumn{2}{c}{NAD} \\ 
    & w/o & w & w/o & w & w/o & w & w/o & w & w/o & w & w/o & w \\  \cline{1-1}  \cline{2-3} \cline{4-5} \cline{6-7} \cline{8-9} \cline{10-11} \cline{12-13}
    MNIST   & 10.65 & 10.62 & 11.38 & 11.06 & 31.63 & {37.36} & {99.41} & 92.30 & 9.87 & 10.19 & 12.56 & {35.53} \\
    CIFAR10 & 14.99 & {73.93} & 11.62 & {39.77} & 21.01 & {32.91}  & 36.32 & {56.08} & 10.40 & {59.70} & 10.66  & {71.69} \\
    GTSRB   & 6.40 & 6.42 & 6.40 & 6.43 &12.60 &10.31 &4.20 & {17.77} &6.43 &6.46 &6.40 &6.56\\
    CelebA  & 29.51 & {100} & 8.43 & {99.99}  & 93.08 & {98.01} & 87.99 & {96.91} & 6.52 & {76.74} & 6.23 & {69.57}\\
    \Xhline{2\arrayrulewidth}
    \end{tabular}
    }
    
    \label{tab:nad_results}
    \vspace{-15pt}
\end{table*}

\subsubsection{Predicting Probability at Different Activated-Backdoor Numbers}
Here, we further investigate the influence of the number of activated backdoors on the predicting probability of the target class at inference time. This study can reveal the degree of over-confidence of the victim model over the backdoors. Figure \ref{fig:probs_by_attack_num} shows the results of this study. As can be seen from the figure, the predicting probabilities of the target class on the poisoned inputs generally increase as the number of activated backdoors increases; in most cases, the predicting probability of the target class is much smaller than 100\%, e.g., about 50\% when one backdoor is activated. In contrast, for conventional dirty-label backdoor attacks, the average predicting probability of the target class is always close to 100\%. This verifies the flexibility of the attacks performed with our strategy. 

\vspace{-5pt}
\subsection{Defense Experiments}
\begin{wrapfigure}{r}{0.5 \textwidth}
\vspace{-5pt}
    \centering
    \includegraphics[width=0.5\textwidth]{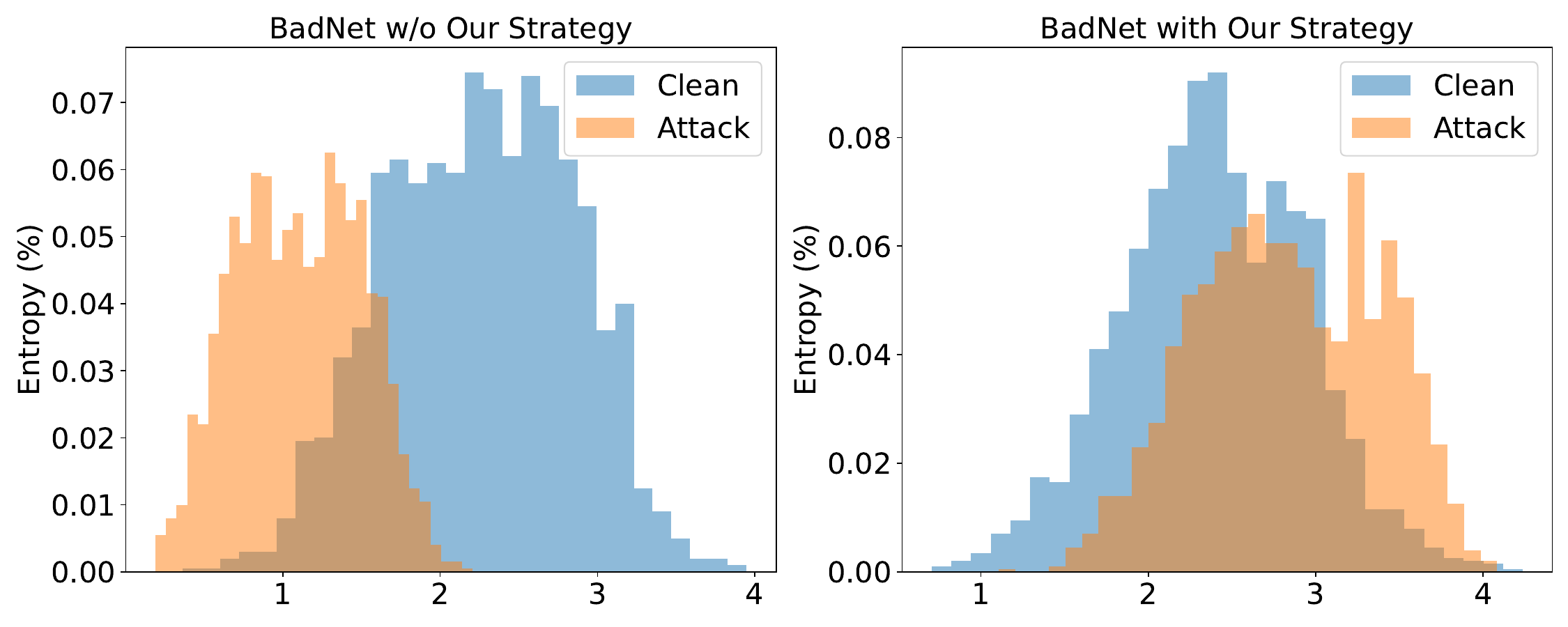}
    \vspace{-10pt}
    \caption{Entropy of the label predicting probabilities on MNIST.}
    \label{fig:strip_mnist}
    \vspace{-15pt}
\end{wrapfigure}
We evaluate the performance of attacks against Fine-Pruning, NAD, STRIP, and Neural-Cleanse. Fine-Pruning and NAD evaluate the stealthiness of attacks via the ASR of the post-processed victim model. STRIP evaluates the stealthiness of attacks by False Acceptance Rate (\textbf{FAR}).  FAR defines the probability that the poisoned input is recognized as the clean input by the defense system. For STRIP, we choose the threshold on the clean test set so that only 1\% of clean samples are recognized as poisoned samples and report the corresponding FAR values of different attacks. For Neural Cleanse, we report the Anomaly Index produced by Neural Cleanse after running it on the poisoned models.


\begin{table}[]
\parbox[t]{.5\textwidth}{ 
\setlength{\tabcolsep}{2pt}

\centering
\scriptsize
\caption{Anomaly Index. A smaller value indicates better stealthiness against Neural-Cleanse. Star means target pattern norm is not the smallest.
    }

\begin{tabular}{lcccccc@{}}
    \toprule
    \multirow{2}{*}{Dataset} & \multicolumn{2}{c}{BadNet} & \multicolumn{2}{c}{SIG} & \multicolumn{2}{c}{WaNet} \\ 
     & w/o & w & w/o & w & w/o & w \\ 
     \midrule
    MNIST  & 2.20 & 1.73$^*$ & 1.87 & 1.19 & 1.57 & 1.26$^*$  \\
    CIFAR10 & 5.75 & 3.93 & 1.63$^*$ & 1.12$^*$ & 0.84$^*$ & 1.81$^*$ \\
    GTSRB & 5.94 & 3.83 & 2.67 & 1.97 & 2.98 & 1.49$^*$ \\
    CelebA & 4.77 & 2.16 & 3.83 & 1.07$^*$ & 1.15$^*$ & 1.31$^*$\\
    \bottomrule
    \end{tabular}
    
    \label{tab:cleanse_results}}
\hfill
\parbox[t]{.47\textwidth}{ 
\setlength{\tabcolsep}{2pt}
\scriptsize
\centering
\caption{False Acceptance Rate (FAR) $\times$ 100. A larger value indicates better stealthiness against STRIP. \\}\vspace{-10pt}
\begin{tabular}{lcccccc@{}}
    \toprule
    \multirow{2}{*}{Dataset} & \multicolumn{2}{c}{BadNet} & \multicolumn{2}{c}{SIG} & \multicolumn{2}{c}{WaNet} \\ 
    & w/o & w  & w/o & w & w/o & w \\ 
    \midrule
    MNIST  & 0.0 & 99.60 & 0.0 & 100 & 1.80 & 100 \\
    CIFAR10 & 97.10 & 100 & 100 & 98.70 &97.00 &99.70\\
    GTSRB & 0.2 & 98.85 & 82.50 & 100 & 68.00 & 100\\
    CelebA & 0.0 & 98.80 & 19.50 & 99.95 & 27.75 &100 \\ 
    \bottomrule
    \end{tabular}
    \label{tab:far_results}}

\end{table}
\noindent \textbf{Fine-Pruning and NAD Defenses.} Table \ref{tab:nad_results} shows the ASRs of different attacks after applying Fine-Pruning and NAD defenses. As can be seen, on CIFAR10 and CelebA, the attacks with our strategy achieve much higher ASRs against Fine-Pruning and NAD than their counterparts without using our strategy. On MNIST and GTSRB, the obtained ASRs of all attacks are small and are close to the ratio of the target class in the test dataset. Our explanation for this phenomenon is that MNIST and GTSRB are two easy-to-fit datasets. Finetuning the model will significantly change the parameters of the model (like training from scratch) while achieving high clean accuracy. This is why the obtained ASRs after the defense are close to the ratio of the target class. All in all, we can see that our strategy can improve the stealthiness of the existing attacks against Fine-Pruning and NAD.


\noindent \textbf{STRIP Defense.} Table \ref{tab:far_results} shows the performance of the attacks against STRIP. As can be seen from the table, attacks with our strategy achieve much higher FAR than their counterparts without using our strategy on MNIST, GTSRB, and CelebA. Surprisingly, on MNIST, all attacks employing our strategy achieve almost 100\% FARs. To explain this observation, we plotted the entropy of the predicting probabilities of the poisoned and clean inputs of MNIST. Figure \ref{fig:strip_mnist} shows the results. We observe that for both attacks, the entropy of many benign inputs is quite small, resulting in a small threshold for rejecting the poisoned inputs and consequently, causing many poisoned inputs to escape the detection of STRIP. In addition, with our strategy, the entropy distribution of the label predicting probabilities of the poisoned inputs is more uniform. This means that it is hard to separate the poisoned from the benign inputs using STRIP.

\noindent \textbf{Neural-Cleanse Defense.} Table \ref{tab:cleanse_results} shows the Anomaly Index values of different attacks. 
A value followed by a star means the pattern norm of the target class is not the smallest one among those of all the classes; that is we will make a mistake if we aim to identify the attacker-defined target class based on the pattern norms.
As can be seen from the table, in most cases, the attack performed with our strategy achieves better stealthiness than its corresponding baseline against Neural-Cleanse. One explanation for the improved stealthiness is that the training loss of the trigger pattern mining consists of the cross-entropy-based label prediction loss and the norm-based regularization loss. For the attack performed with our strategy, the predicted probability of the target class will be considerably smaller than 100\% even if the pattern is the backdoor trigger of the attack. In this case, the regularization loss is small, but the label prediction loss will be large. Thus, it may increase the size of the pattern for a better balance between the label prediction loss and the regularization loss. 

\begin{table}[]
\parbox[t]{.47\linewidth}{ 
\scriptsize
\centering
\caption{The Elimination Rate (ER) and Sacrifice Rate (SR) of outlier detection on MNIST.}
\vspace{-3pt}
\mbox{

    \renewcommand{\arraystretch}{0.3}
    
\begin{tabular}{llcccccc}
\toprule
\multicolumn{2}{c}{\multirow{2}{*}{Attack}}  & \multicolumn{2}{c}{SPECTRE} & \multicolumn{2}{c}{SS} \\ 
 \cmidrule(lr){3-4}\cmidrule(lr){5-6}
          & &  ER    & SR    & ER    & SR    \\ \midrule
\multirow{2}{*}{BadNet} & w   &  22.71 & 14.79 & 49.01 & 14.09 \\
                        & w/o &  99.34 & 14.13 & 100   & 14.13 \\ \midrule
\multirow{2}{*}{SIG}    & w   &  57.60 & 13.86 & 53.94 & 13.96 \\
                        & w/o &  99.51 & 14.13 & 100   & 14.13 \\ \midrule
\multirow{2}{*}{WaNet}  & w   &  19.56 & 14.87 & 45.99 & 14.17 \\
                        & w/o &  95.41 & 14.17 & 94.75 & 14.18 \\ \bottomrule
\end{tabular}
}
\label{tab:detection_rate}
}
\hfill
\parbox[t]{.47\linewidth}{
\scriptsize
\centering
\caption{The accuracy and ASR after eliminating suspicious samples to train on MNIST.}
\vspace{-3pt}
\mbox{
\renewcommand{\arraystretch}{0.3}
\begin{tabular}{llrrrrr}
\toprule
\multicolumn{2}{c}{\multirow{2}{*}{Attack}} &
  \multicolumn{2}{c}{SPECTRE} &
  \multicolumn{2}{c}{SS} \\ \cmidrule(lr){3-4}\cmidrule(lr){5-6}
& & ACC & ASR & ACC & ASR \\ \midrule
\multirow{2}{*}{BadNet} & w   &  99.29 & 100   & 99.26 & 100   \\
                        & w/o &  99.27 & 10.37 & 99.29 & 10.40  \\ \midrule
\multirow{2}{*}{SIG}    & w   & 98.96 & 52.36 & 99.02 & 63.31 \\
                        & w/o &  99.32 & 10.25 & 99.19 & 10.39 \\ \midrule
\multirow{2}{*}{WaNet}  & w   &  99.09 & 93.57 & 99.20  & 81.32 \\
                        & w/o &  99.01 & 10.42 & 99.05 & 10.42 \\ \bottomrule
\end{tabular}
}
\label{tab:retraining_defense}
}
\vspace{-19pt}
\end{table}
\noindent \textbf{Outlier Removal Defense.} We also evaluate the stealthiness and resilience of our strategy against outlier detection in which the defense examines the training dataset and filters out suspicious samples. Table~\ref{tab:detection_rate} illustrates the Elimination Rate (ER) and Sacrifice Rate (SR), the ratio of backdoor samples detected and the ratio of clean samples misidentified, of SPECTRE~\cite{hayase2021spectre} and Spectral Signature (SS)~\cite{tran2018spectral} on MNIST. For the defense to be effective, it must have both high ER and low SR. As can be observed, our strategy substantially decreases the ER of outlier detection methods, making the attack resilient to the defense. To further facilitate, we train a model with a new dataset in which suspicious samples are eliminated. Table~\ref{tab:retraining_defense} shows that after the defense is applied, our strategy helps the attack keep a high poisoning rate, while the baseline method fails in manipulating the model. 
\vspace{-10pt}
\section{Conclusion}
\vspace{-6pt}

This work introduces a novel strategy to mitigate the over-confidence problem of dirty-label backdoor attacks. The idea of our strategy originates from label smoothing and changes the probability of modifying the label of a poisoned sample to the target class from 100\% to an input-specific value of $p_t(\bm{x})$. Empirical studies on three popular types of attacks and four prevalent defense algorithms show that our strategy can effectively alleviate the over-confidence problem of dirty-label backdoor attacks and accordingly, improve their stealthiness against the defense algorithms that build on the over-confidence characteristic of backdoor attacks. This urges backdoor researchers to devise defensive measures to counter this type of attack.
\clearpage  

%
%
\bibliographystyle{plainnat}
\bibliography{ref}
\newpage
\appendix
\input{supp}

\end{document}

%% file: supp.tex
This document provides additional details, analysis, and experimental results to support the main submission. We begin by providing the proof of Theorem 3.1 in Section~\ref{sec:proof}. Then we provide additional defense experiments in Section~\ref{sec:more_defense}. Finally, we include more attack experiments in Section~\ref{sec:more_attacks}, demonstrating the effectiveness of our method.
\section{Proof of Theorem 1}\label{sec:proof1}
\begin{proof}
Let $y_s, y_t \in \{0, 1\}$ denote two indicators with $y_s=1$ and $y_t=1$ indicating that $\bm{x}$ and $\tilde{\bm{x}}$ belongs to the target class, respectively. For expression simplicity and without confusion, denote $p(y=c|\bm{x}; \mathcal{D})$ as $p(y=c|\bm{x})$ and $p(y=c|\tilde{\bm{x}}; \tilde{\mathcal{D}})$ as $p\left(y=c|\tilde{\bm{x}}\right)$.
Then, we have:


\begin{align*} 
    \hat{\beta}({\bm{x}}) &= p(y_t=1| \tilde{\bm{x}}) = \sum_{c=0}^1 p\left(y_t=1, y_s=c|\tilde{\bm{x}}\right) \\
    &= \sum_{c=0}^1 p(y_t=1|y_s=c, \tilde{\bm{x}})p(y_s=c| \tilde{\bm{x}}).
\end{align*}
For most backdoor injection functions, it is reasonable to assume $\mathcal{B}(\cdot)$ is an injective function, thus:
\begin{align*}
\begin{split}
    \hat{\beta}({\bm{x}}) = \sum_{c=0}^1 p\left(y_t=1|y_s=c, \tilde{\bm{x}}\right)p(y_s=c|\bm{x})
    \end{split}
\end{align*}
According to the poisoning process defined in Eq. (\ref{eq:poison}), $y_s=1 \rightarrow y_t=1$ and $\mathbb{E}_{\tilde{\mathcal{D}}}\left[p\left(y_t=1|y_s=0, \tilde{\bm{x}}\right)\right] = p_t(\bm{x})$. Therefore, we have:
\begin{align*}
    \mathbb{E}[\hat{{\beta}}({\bm{x}})] & = p_t(\bm{x})p(y_s=0| \bm{x}) + p(y_s=1|\bm{x}) \\
    &= \alpha(\bm{x}) + (1-\alpha(\bm{x}))p_t(\bm{x}),
\end{align*}
completing the proof.
\end{proof}
\section{Proof of Theorem 2}\label{sec:proof}

\begin{proof}
According to Theorem 1 in the main manuscript, we can easily get that:
\begin{align}
    \hat{\beta}_i^\prime & = \sum_{c=0}^1 p\left(y_t=1|y_s=c, \bm{\zeta}(\bm{x}_i)\right) p(y_s=c|\bm{x}_i)\\
    & = p\left(y_t=1|y_s=0, \bm{\zeta}(\bm{x}_i)\right) (1-\alpha_i) + \alpha_i 
\end{align}
According to the data poison process of LBSAs, the expected value of $p\left(y_t=0|y_s=0, \bm{\zeta}(\bm{x}_i)\right)$ is the probability that all the $k$ attacks do not change the label of $\bm{x}$ to the target class, i.e.,:
\begin{align}
    \mathbb{E}_{\tilde{\mathcal{D}}}\left[ p\left(y_t=0|y_s=0, \bm{\zeta}(\bm{x}_i)\right)\right] = (1-p_n(\bm{x}_i))^k.
\end{align}
Therefore, 
\begin{align}
    \mathbb{E}[\hat{\beta}_i^\prime] =  (1-\alpha_i) \left[1- (1-p_n(\bm{x}_i))^k\right] + \alpha_i > \beta_i,
\end{align}
completing the proof.
\end{proof}

\section{More Defense Experiments}\label{sec:more_defense}
In this section, we evaluate our approach against additional defense.
Note that our evaluation covers a broader range of defense methods, including pruning-based, input-perturbation, fine-tuning, and reversed-trigger engineering. Nevertheless, we acknowledge that there exist other defenses in each category that we have not tested our defenses against. For example, \citet{shen2021backdoor} primarily addresses the high computational complexity of Neural Cleanse and shares the same reverse-trigger engineering principle and approach to Neural Cleanse. 

\subsection{Performance against Adversarial Unlearning of Backdoors}

We evaluate our method against I-BAU~\cite{zeng2021adversarial}, which falls into a similar class of defenses as NAD. The results based on WaNet after the defense of I-BAU are provided in Table~\ref{tab:defense_ibau}.

\begin{table}[h!]
    \centering
    \caption{Performance against I-BAU~\cite{zeng2021adversarial}}
    \begin{tabular}{c|c|cc}
        \toprule
         Dataset	& Method	& Acc & 	ASR\\\hline
         
        CIFAR10	& w/o	& 86.10& 	7.03\\
CIFAR10 & 	w& 	87.68& 	82.95 \\
GTSRB& 	w/o& 	99.02& 	5.40\\
GTSRB& 	w	& 99.72& 	17.28\\
        \bottomrule

    \end{tabular}
    
    \label{tab:defense_ibau}
\end{table}

As we can observe, similar to the results on NAD, I-BAU is not effective against our attack. Since our proposed strategy aims to make the poisoned samples be predicted as the target class with a probability slightly greater than that of other classes, while still remaining below 0.6, the trigger is not effective for increasing the classification loss in I-BAU's maximization step. Thus, our attack can evade the defense of I-BAU.

\subsection{Entropy Distribution in STRIP}

\begin{figure}[h!]
    \centering     
    \includegraphics[width=0.5\textwidth]{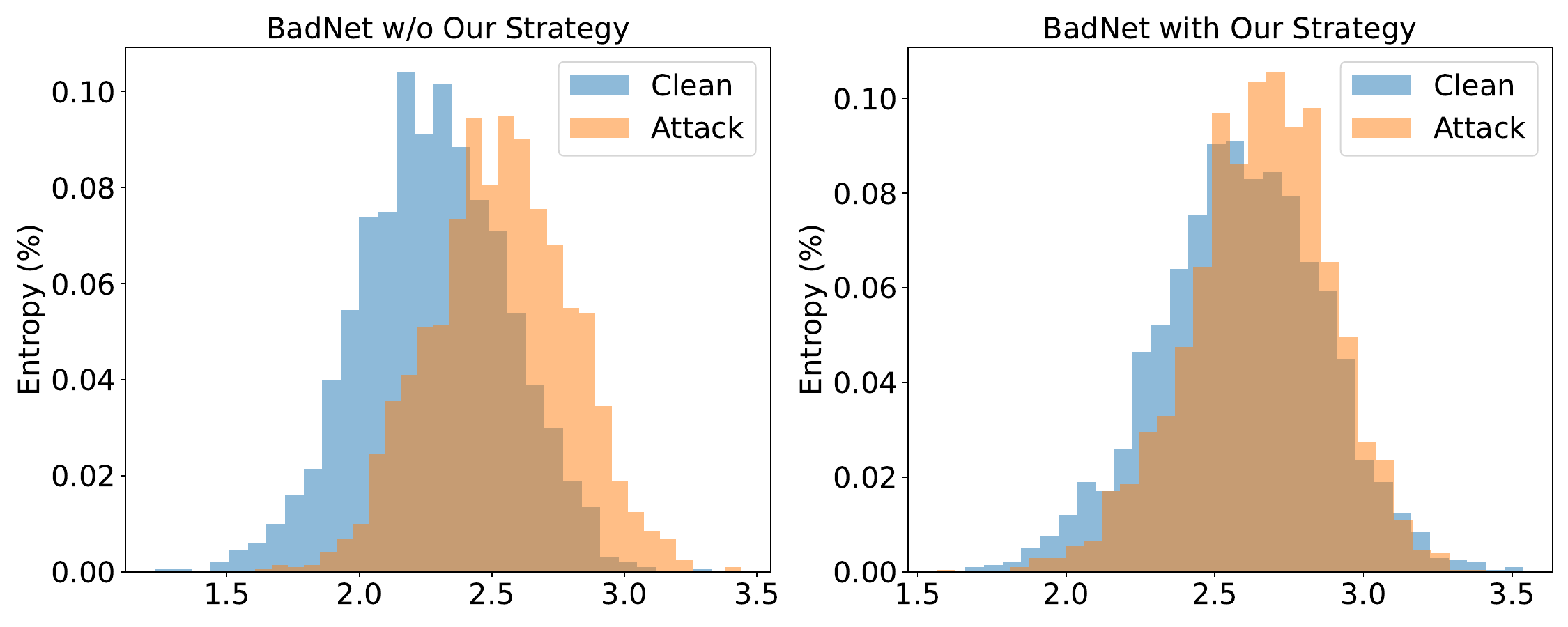}
    \caption{Entropy distribution on CIFAR10 in STRIP.}
    \label{fig:strip_on_cifar}
\end{figure}

Figure~\ref{fig:strip_on_cifar} shows the entropy distribution of our method and the hard attacks on CIFAR10 in STRIP. From these figures, we can see that our label-smoothing strategy can indeed improve the stealthiness of the hard attacks in front of STRIP. 


\subsection{Average Entropy \& Variance Distribution of NLS in STRIP}

\begin{figure}[h!]
    \centering    
    \includegraphics[width=0.235\textwidth]{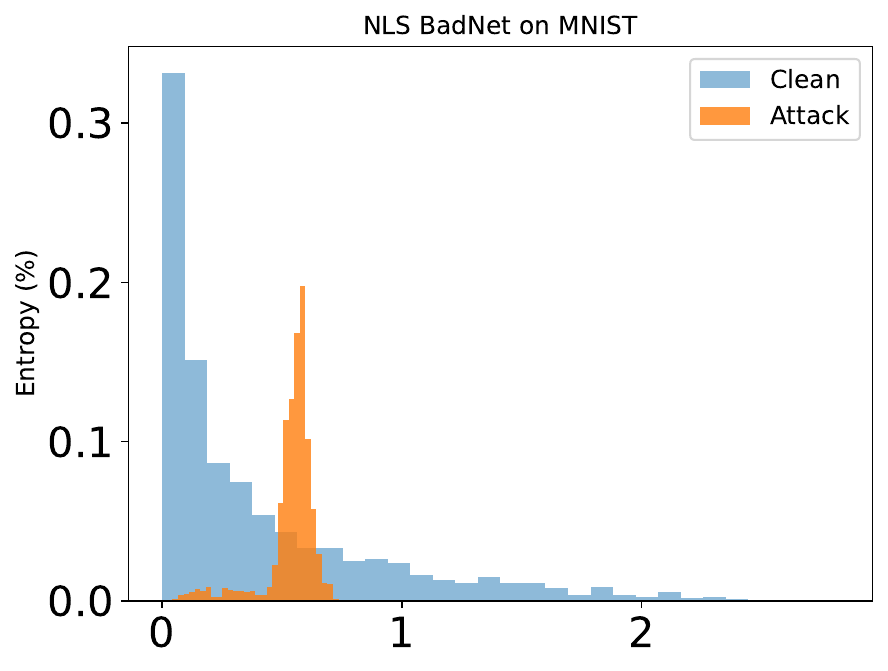} 
    \includegraphics[width=0.235\textwidth]{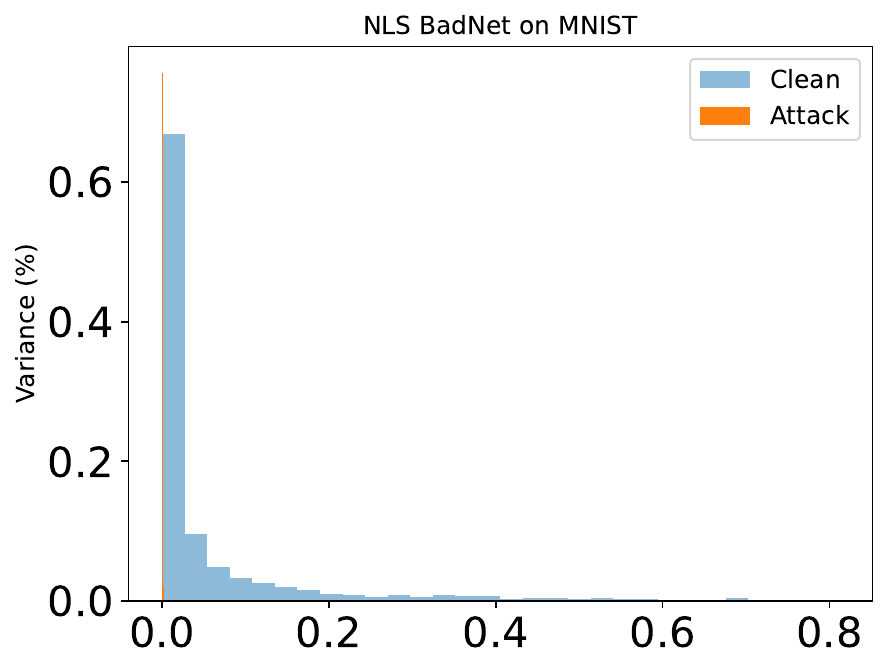}
    \caption{Distributions of the average entropy and variance for the BadNet-based attack on MNIST, which applies NLS.}
    \label{fig:mnist_badnet_nls}
\end{figure}

\begin{figure}[h!]
    \centering    
    \includegraphics[width=0.235\textwidth]{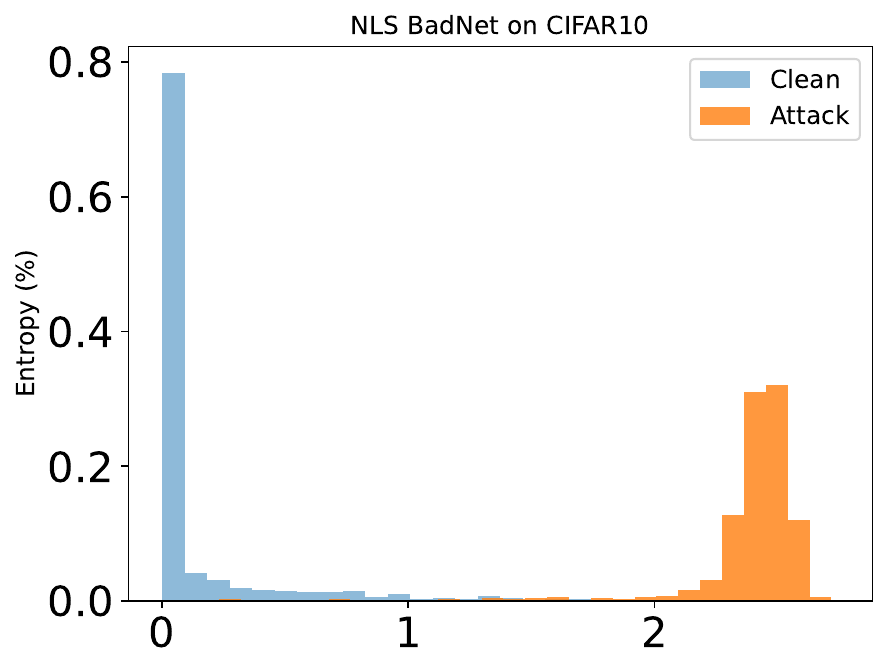} 
    \includegraphics[width=0.235\textwidth]{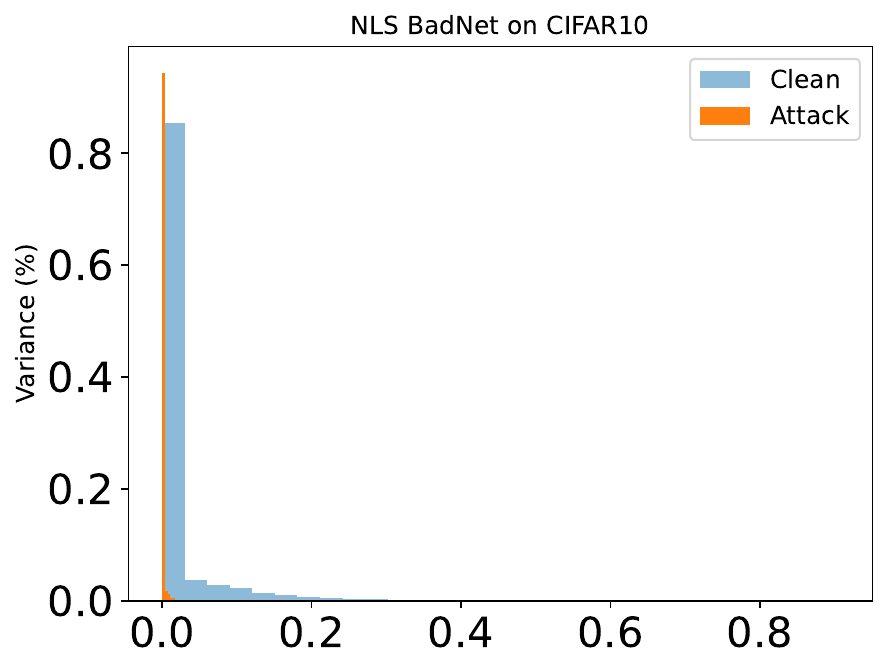}
    \caption{Distributions of the average entropy and variance for the BadNet-based attack on CIFAR10, which applies NLS.}
    \label{fig:cifar10_badnet_nls}
\end{figure}

\begin{figure}[h!]
    \centering    
    \includegraphics[width=0.235\textwidth]{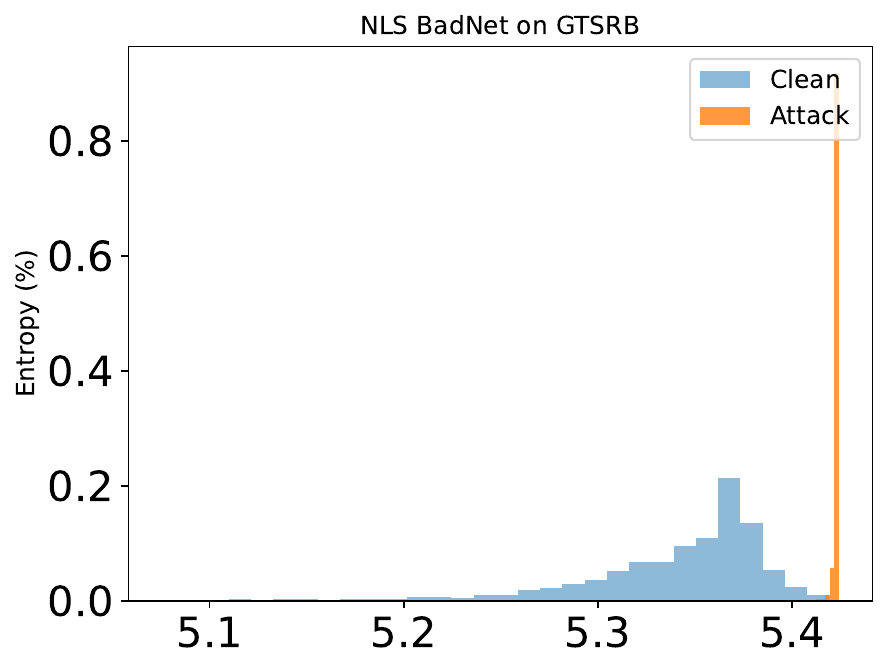}
    \includegraphics[width=0.235\textwidth]{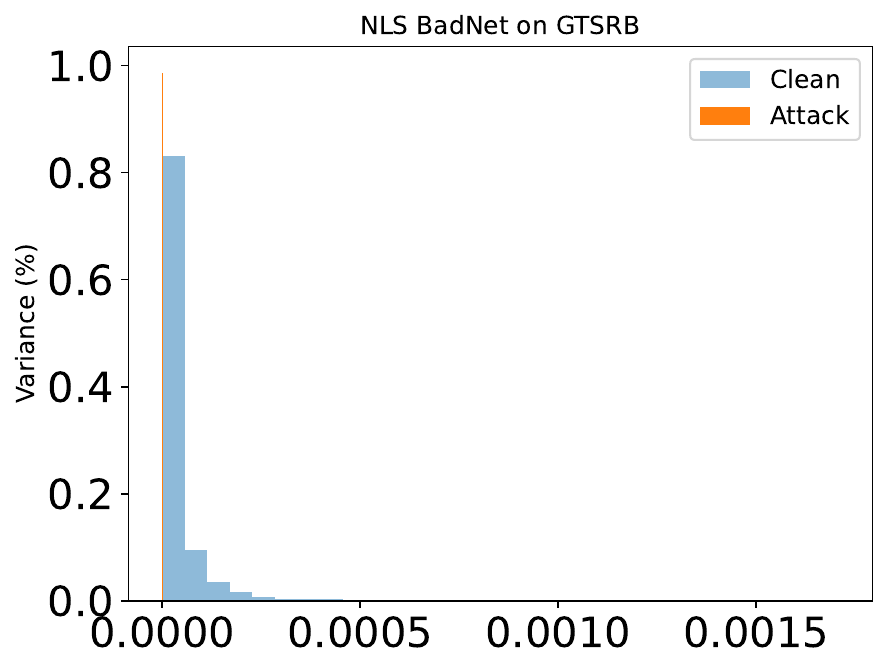}
    \caption{Distributions of the average entropy and variance for the BadNet-based attack on GTSRB, which applies NLS.}
    \label{fig:gtsrb_badnet_nls}
\end{figure}

\begin{figure}[h!]
    \centering    
    \includegraphics[width=0.235\textwidth]{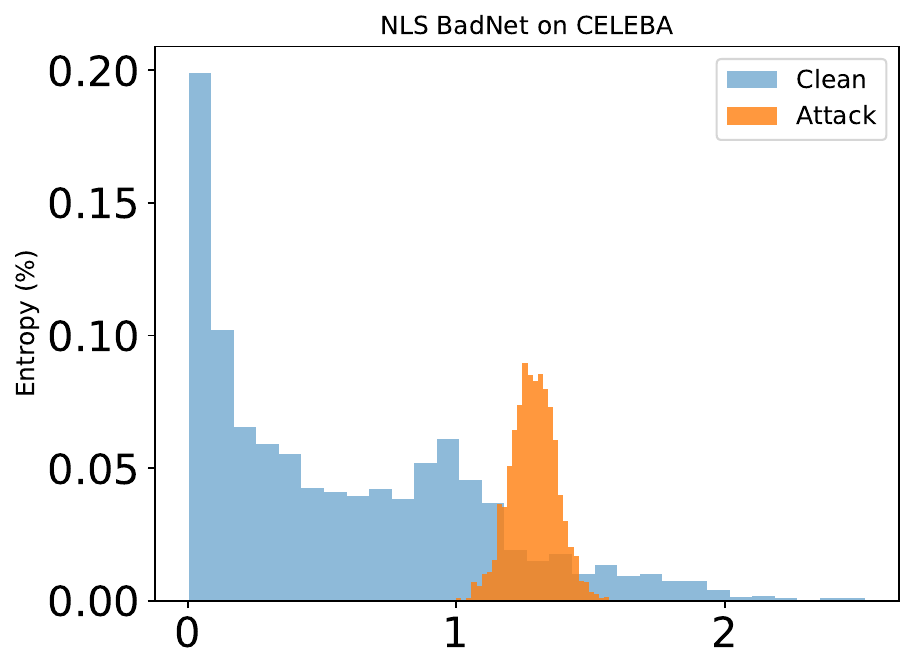}
    \includegraphics[width=0.235\textwidth]{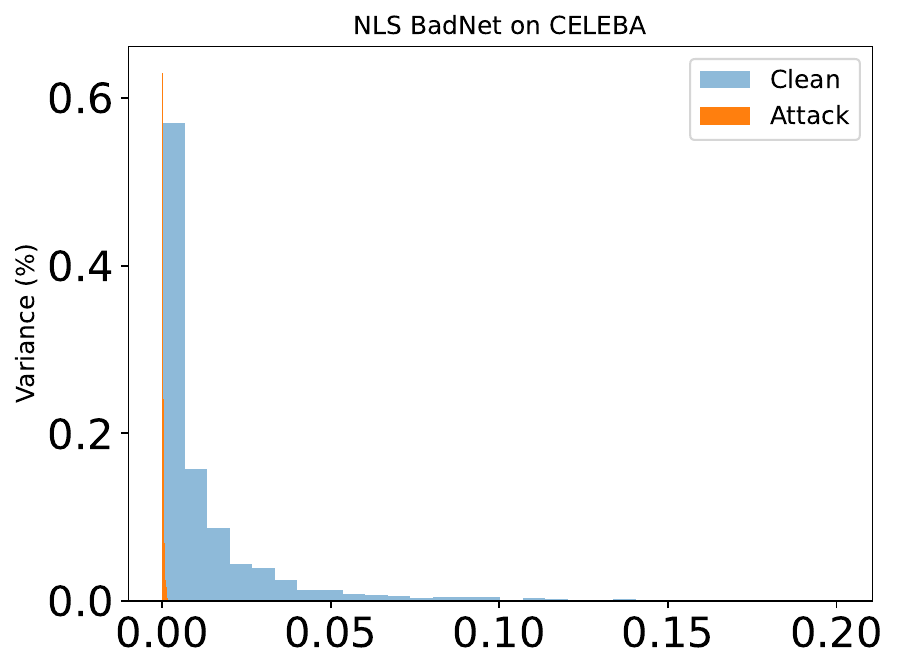}
    \caption{Distributions of the average entropy and variance for the BadNet-based attack on CelebA, which applies NLS.}
    \label{fig:celeba_badnet_nls}
\end{figure}

\begin{figure}[h!]
    \centering    
    \includegraphics[width=0.235\textwidth]{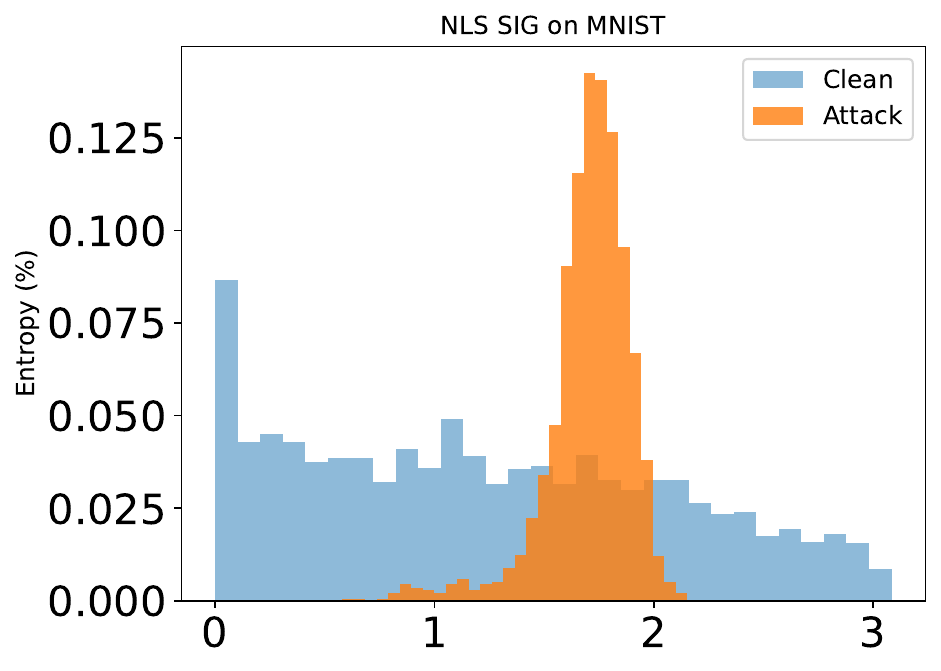}
    \includegraphics[width=0.235\textwidth]{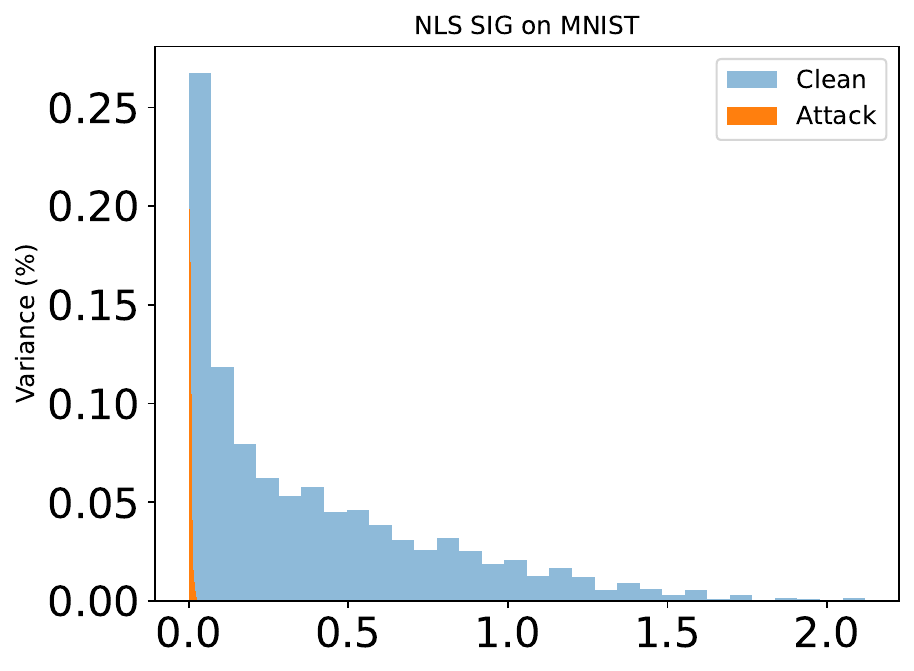}
    \caption{Distributions of the average entropy and variance for the SIG-based attack on MNIST, which applies NLS.}
    \label{fig:mnist_sig_nls}
\end{figure}

\begin{figure}[h!]
    \centering    
    \includegraphics[width=0.235\textwidth]{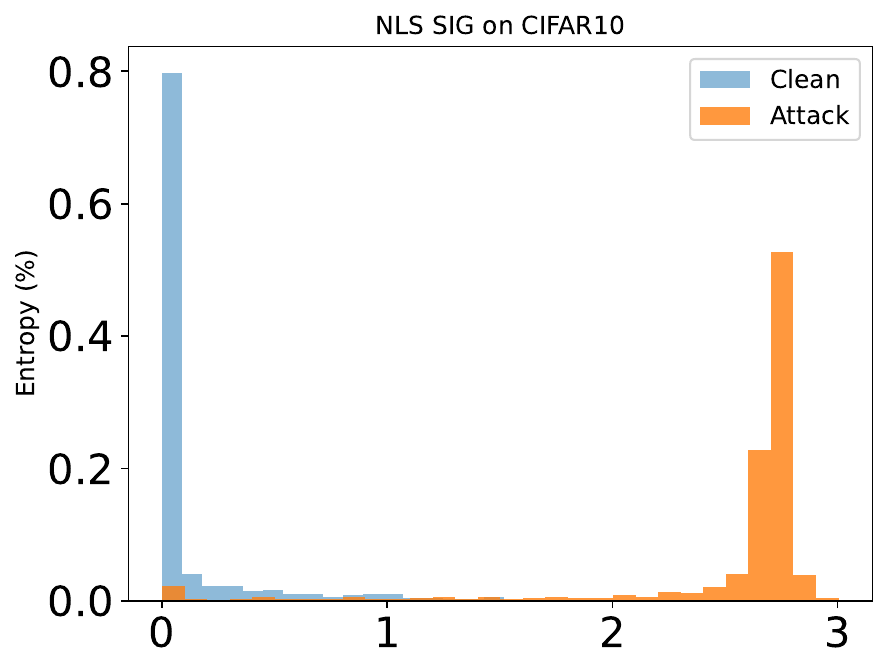} 
    \includegraphics[width=0.235\textwidth]{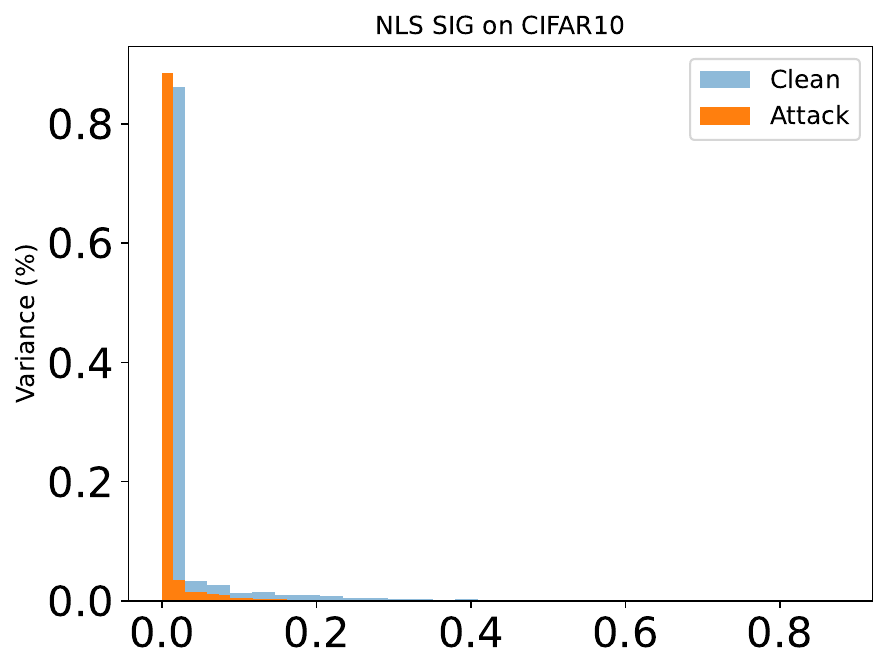}
    \caption{Distributions of the average entropy and variance for the SIG-based attack on CIFAR10, which applies NLS.}
    \label{fig:cifar10_sig_nls}
\end{figure}

\begin{figure}[h!]
    \centering    
    \includegraphics[width=0.235\textwidth]{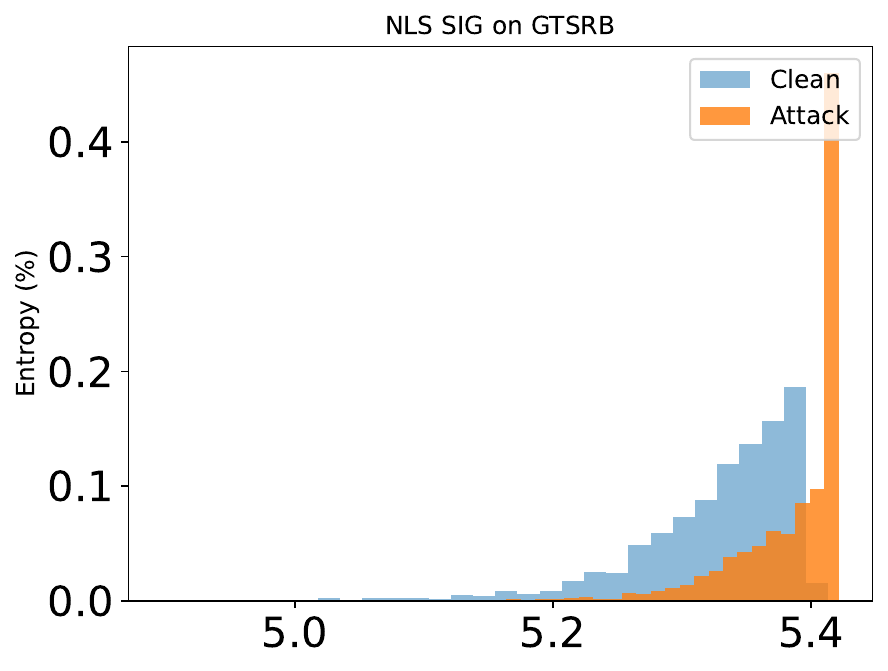} 
    \includegraphics[width=0.235\textwidth]{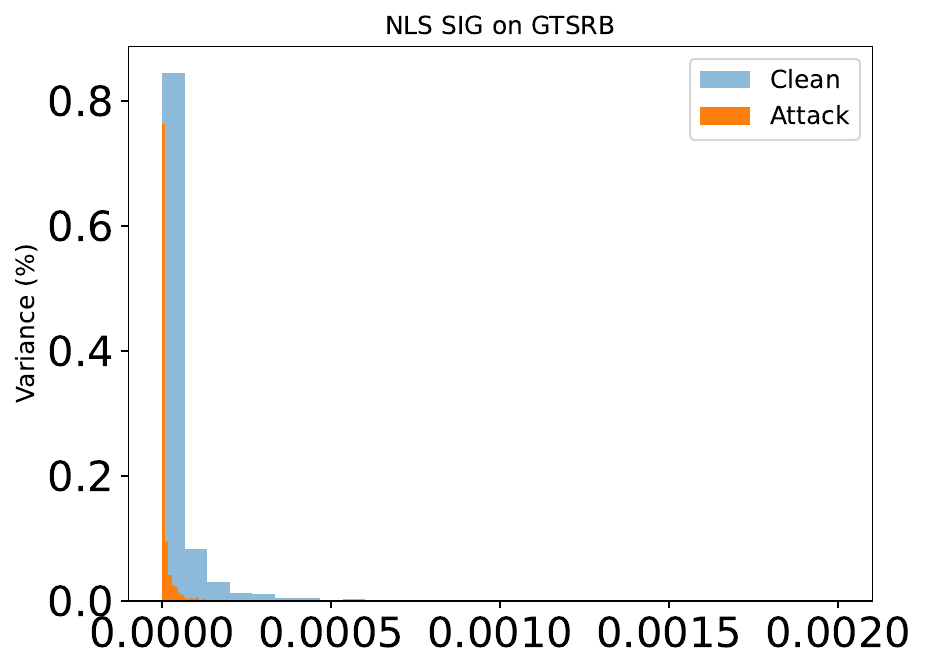}
    \caption{Distributions of the average entropy and variance for the SIG-based attack on GTSRB, which applies NLS.}
    \label{fig:gtsrb_sig_nls}
\end{figure}

\begin{figure}[h!]
    \centering    
    \includegraphics[width=0.235\textwidth]{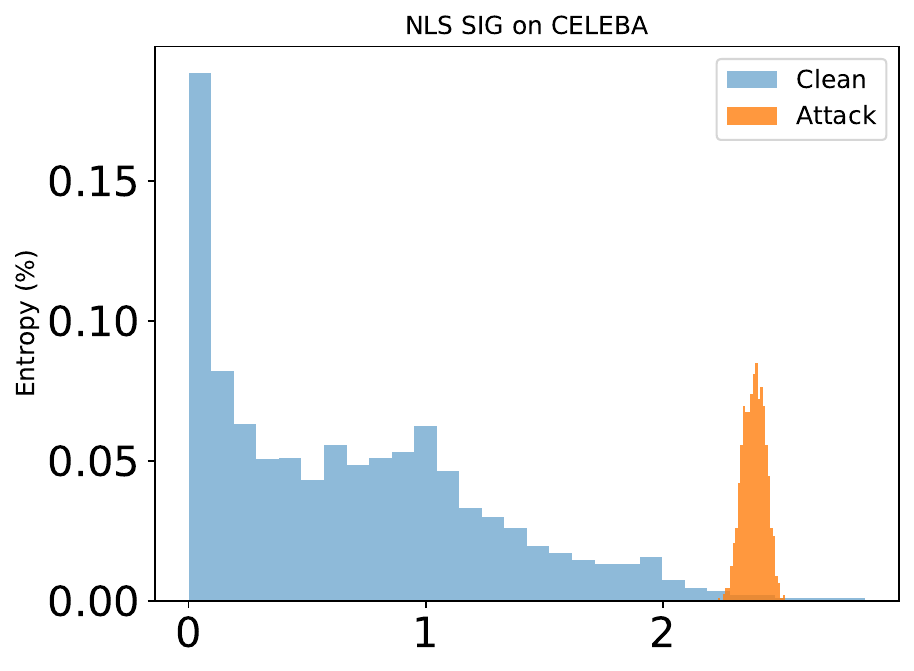} 
    \includegraphics[width=0.235\textwidth]{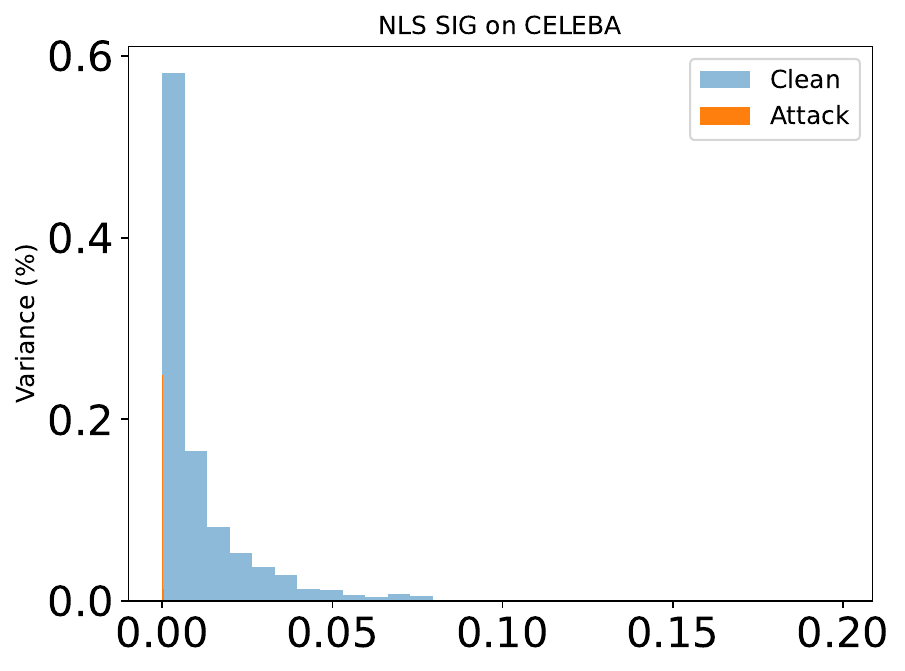}
    \caption{Distributions of the average entropy and variance for the SIG-based attack on CelebA, which applies NLS.}
    \label{fig:celeba_sig_nls}
\end{figure}

\begin{figure}[h!]
    \centering    
    \includegraphics[width=0.235\textwidth]{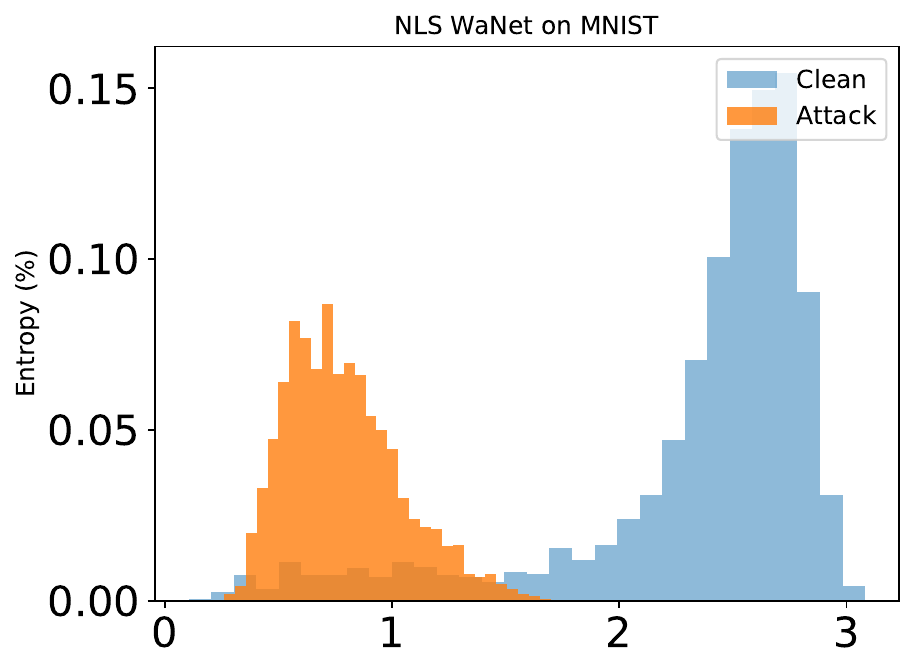}
    \includegraphics[width=0.235\textwidth]{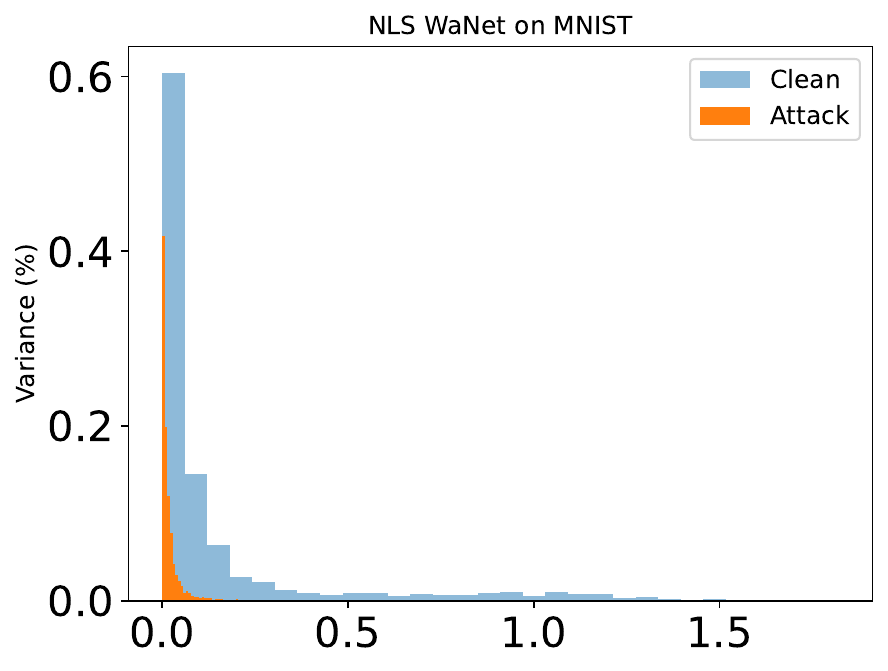}
    \caption{Distributions of the average entropy and variance for the WaNet-based attack on MNIST, which applies NLS.}
    \label{fig:mnist_wanet_nls}
\end{figure}

\begin{figure}[h!]
    \centering    
    \includegraphics[width=0.235\textwidth]{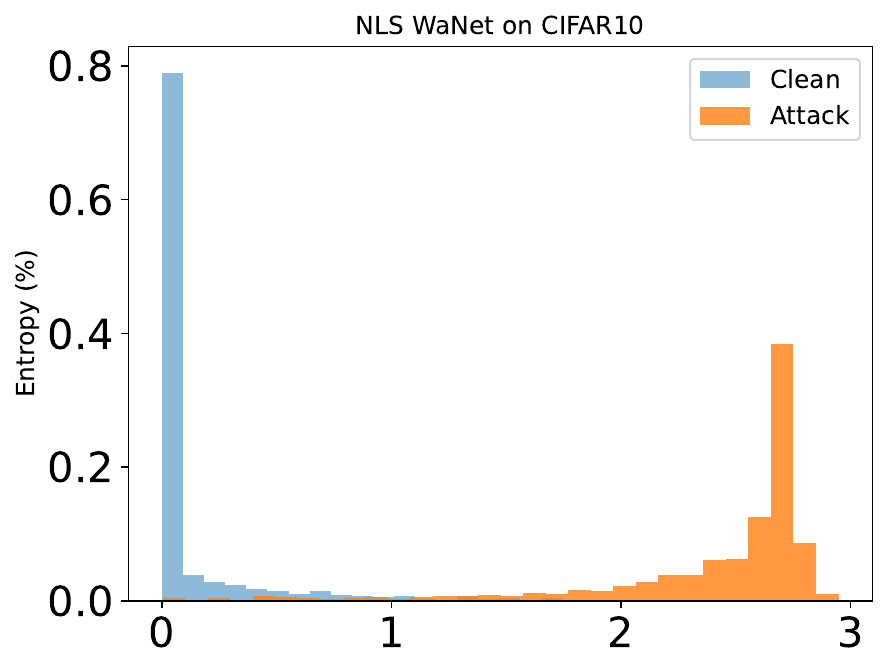}
    \includegraphics[width=0.235\textwidth]{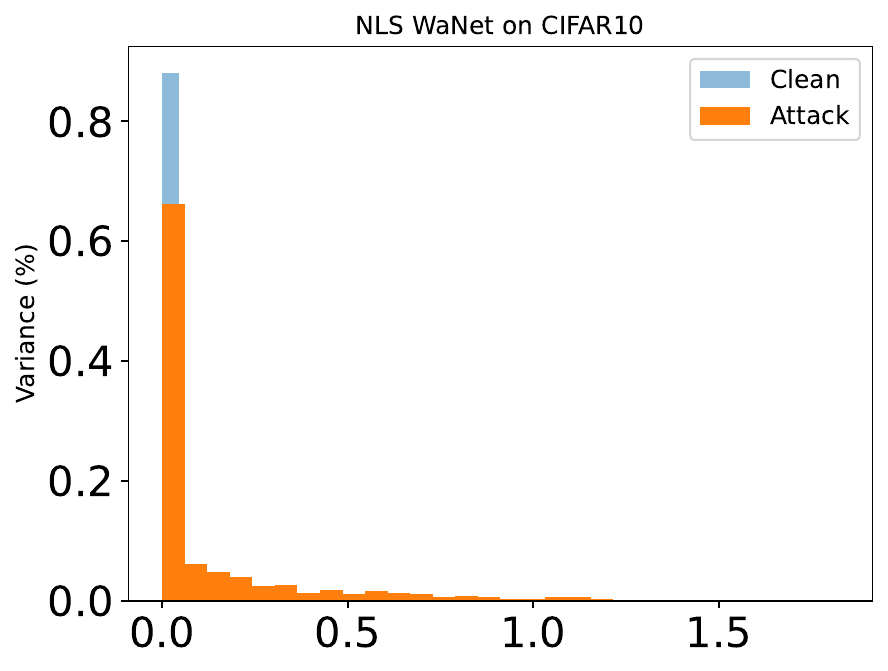}
    \caption{Distributions of the average entropy and variance for the WaNet-based attack on CIFAR10, which applies NLS.}
    \label{fig:cifar10_wanet_nls}
\end{figure}

\begin{figure}[h!]
    \centering    
    \includegraphics[width=0.235\textwidth]{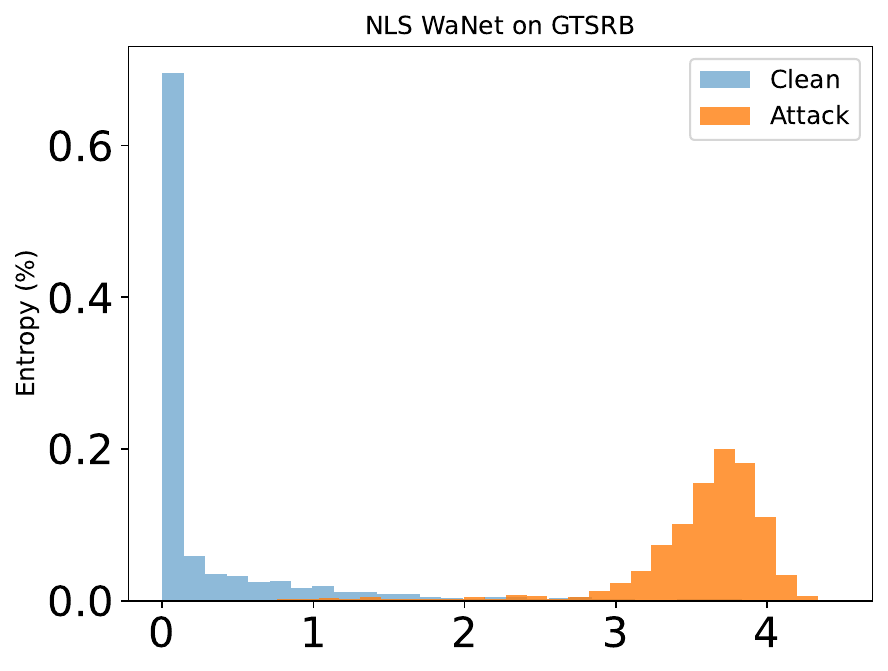}
    \includegraphics[width=0.235\textwidth]{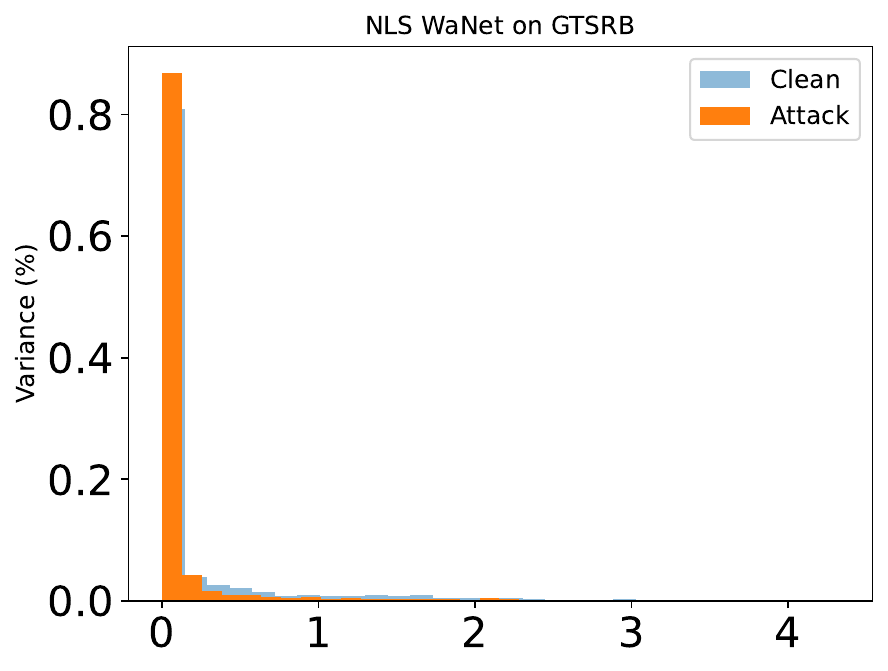}
    \caption{Distributions of the average entropy and variance for the WaNet-based attack on GTSRB, which applies NLS.}
    \label{fig:gtsrb_wanet_nls}
\end{figure}

\begin{figure}[h!]
    \centering    
    \includegraphics[width=0.235\textwidth]{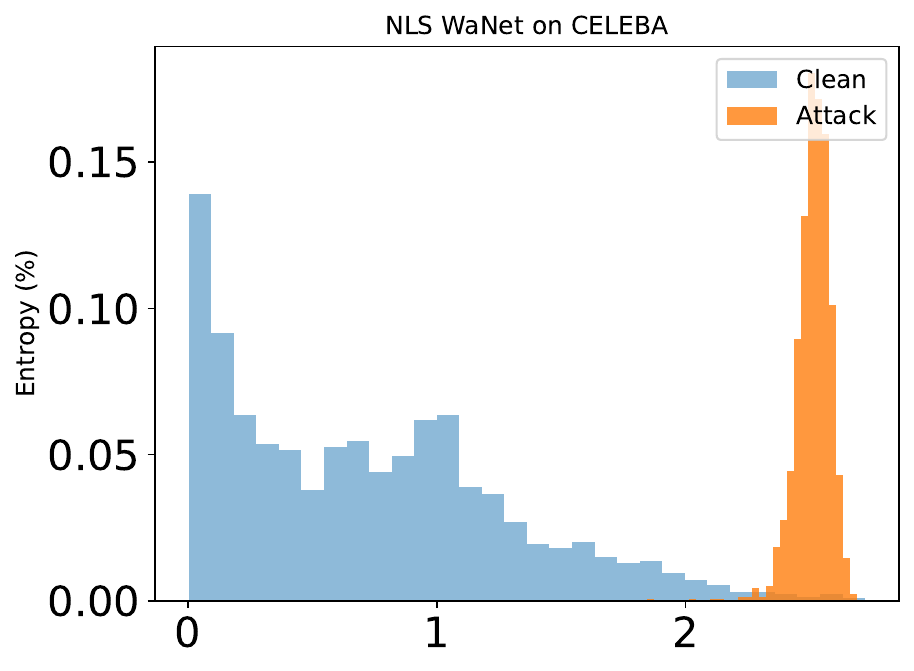}
    \includegraphics[width=0.235\textwidth]{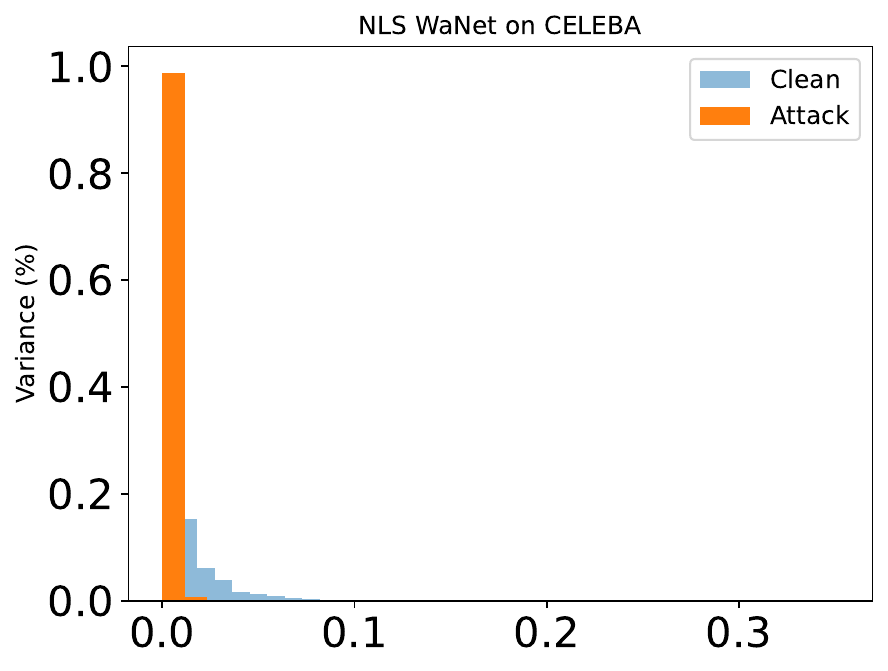}
    \caption{Distributions of the average entropy and variance for the WaNet-based attack on CelebA, which applies NLS.}
    \label{fig:celeba_wanet_nls}
\end{figure}

Figures \ref{fig:mnist_badnet_nls}, \ref{fig:cifar10_badnet_nls}, \ref{fig:gtsrb_badnet_nls}, \ref{fig:celeba_badnet_nls},  \ref{fig:mnist_sig_nls}, \ref{fig:cifar10_sig_nls}, \ref{fig:gtsrb_sig_nls}, \ref{fig:celeba_sig_nls}, \ref{fig:mnist_wanet_nls}, \ref{fig:cifar10_wanet_nls}, \ref{fig:gtsrb_wanet_nls}, and \ref{fig:celeba_wanet_nls} shows the average entropy and variance distribution of our method that applies the NLS strategy.

\subsection{Outlier Detection Defenses}
In this section, we provide the results of outlier detection defenses, including Spectral Signature and SPECTRE. Table~\ref{tab:detection_rate1} presents the elimination rate and sacrifice rate on CIFAR10, showing that our method helps the model stay stealthy under these defensive strategies. 
\begin{table}[ht!]
\scriptsize
\centering
\caption{The Elimination Rate (ER) and Sacrifice Rate (SR) of outlier detection on CIFAR10.}
\begin{tabular}{llcccccc}
\toprule
\multicolumn{2}{c}{\multirow{2}{*}{Attack}}  & \multicolumn{2}{c}{SPECTRE} & \multicolumn{2}{c}{SS} \\ 
 \cmidrule(lr){3-4}\cmidrule(lr){5-6}
          & &  ER    & SR    & ER    & SR    \\ \midrule
\multirow{2}{*}{BadNet} & w   & 57.98	& 13.88	& 57.11	& 13.98  \\
                        & w/o &  69.00	& 14.45 &	67.00	& 14.47 \\ \midrule
\multirow{2}{*}{SIG}    & w   &  57.82	& 13.89	& 53.00	& 14.01 \\
                        & w/o &  100& 	14.14 &	100 &	14.14 \\ \midrule
\multirow{2}{*}{WaNet}  & w   &  38.00 &	14.40 &	35.47 &	14.47 \\
                        & w/o &  99.00 &	14.15 &	99.20 &	14.15 \\ \bottomrule
\end{tabular}

\label{tab:detection_rate1}
\end{table}
\section{More Attack Experiments}\label{sec:more_attacks}

\subsection{Estimate $alpha(x)$ with Transformer-based Model}
The table below shows the average ASRs (times 100) of our method when using a transformer-based model (ViT) or a CNN-based model to implement $f_{clean}$ (1-4 denotes the activated backdoor number):
\begin{table}[h!]
    \centering
    \caption{Performance when estimating $alpha(x)$ with Transformer-based Model}
    \begin{tabular}{c|c|cccc}
        \toprule
         Dataset	& $f_{clean}$ &	1	 &	2 &		3	 &	4 \\ \hline
CIFAR10	 &	Transformer &		89	 &	100 &	100	 &	100 \\
CIFAR10	 &	CNN	 &	91	 &	99	 &	100	 &	100 \\
GTSRB	 &	Transformer &		62 &		97 &		99 &		100 \\
GTSRB	 &	CNN  &		60 &		95 &		99 &		99 \\
        \bottomrule

    \end{tabular}
    
    \label{tab:_ibau}
\end{table}

As can be observed, there is minimal variation in the results when using different architectures for $f_{clean}$. While having an accurate value of $\alpha(x)$ can enhance the attack efficiency, it is not essential for a successful attack. In practical scenarios where the estimator of $\alpha(x)$ may not be so reliable, increasing the value of $\beta(x)$ (defined in Eq. (9)) slightly can mitigate its influence (e.g., $\beta(x) = \alpha(x) + 0.2$).

\subsection{Attack with Uniform Probability}

We have introduced a variant of our method which sets $p_t(x) = c$, and it changes the same number of samples' labels as our original method. We report ASRs of this variant and our method based on the invisible WaNet (Consistent refers to the variant, Our denotes our solution, and 1-4 denotes the activated backdoors) in Table~\ref{tab:attack_uniform}.

\begin{table}[h!]
    \centering
    \caption{Attack Performance with uniform $p_t(x)$.}
    \begin{tabular}{l|l|cccc}
        \toprule
Dataset & Solution & 1 & 2 & 3 & 4 \\ \hline
MNIST & Consistent & 83.26 & 96.54 & 99.27 & 99.99 \\
MNIST & Our & 88.80 & 99.589 & 99.99 & 100.00 \\
CIFAR10 & Consistent & 78.32 & 89.92 & 97.28 & 98.09 \\
CIFAR10 & Our & 91.48 & 99.84 & 100.00 & 100.00 \\
GTSRB & Consistent & 43.15 & 68.78 & 87.54 & 97.29 \\
GTSRB & Our & 60.58 & 95.67 & 99.08 & 99.64 \\
CelebA & Consistent & 69.32 & 84.88 & 95.67 & 97.33 \\
CelebA & Our & 76.99 & 98.51 & 99.65 & 99.77 \\
\bottomrule
\end{tabular}
    
    \label{tab:attack_uniform}
\end{table}

As can be observed, our method outperforms the consistent variant, particularly when the number of activated backdoors is small (e.g., 1 or 2). This is not surprising, given that a dynamic value of $p_t(x)$ is more effective for attacking, as demonstrated in Theorem 1. Such a dynamic value places more data poisoning budget on samples located farther from the classification boundary, which is more effective for the attack.